\documentclass[12pt,preprint]{aastex}
\DeclareMathAlphabet{\pazocal}{OMS}{zplm}{m}{n}
\usepackage{multirow}
\usepackage{hhline}
\usepackage{times}
\usepackage{graphicx}
\usepackage[T1]{fontenc}
\usepackage{aecompl} 
\usepackage{amsmath}               
\usepackage{placeins}
\usepackage{booktabs}
\usepackage{amsfonts}               % American Mathematical Society fonts
\usepackage{amssymb}                % American Mathematical Society symbol
\usepackage{rotating}
\usepackage{color}
\usepackage{url}
\usepackage{hyperref}
\usepackage{aas_macros}
\usepackage{xcolor}
\usepackage{chapterbib}
\usepackage{lineno}
\DeclareOldFontCommand{\rm}{\textrm}{\mathrm}

\begin{document}
\title{Uranus and Neptune as methane planets: producing icy giants from refractory planetesimals}
	
\author{Uri Malamud$^{1}$, Morris Podolak$^{2}$, Joshua I. Podolak$^{3}$, Peter H. Bodenheimer$^{4}$}

\affil{$^1$Physics Department, Technion - Israel institute of Technology, Haifa,
	Israel\\$^2$School of the Environment and Earth Sciences, Tel Aviv University, Ramat Aviv, Tel Aviv, Israel\\$^3$D. E. Shaw Group, New York, NY\\$^4$Astronomy and Astrophysics Department, University of California Santa Cruz, CA 95064, USA}

\maketitle	 
	
\section*{Abstract}
Uranus and Neptune are commonly considered ice giants, and it is often assumed that, in addition to a solar mix of hydrogen and helium, they contain roughly twice as much water as rock. This classical picture has led to successful models of their internal structure and has been understood to be compatible with the composition of the solar nebula during their formation \citep{ReynoldsSummers-1965,PodolakCameron-1974,PodolakReynolds-1984,PodolakEtAl-1995,NettelmannEtAl-2013}. However, the dominance of water has been recently questioned \citep{TeanbyEtAl-2020,HelledFortney-2020,PodolakEtAl-2022}. Planetesimals in the outer solar system are composed mainly of refractory materials, leading to an inconsistency between the icy composition of Uranus and Neptune and the ice-poor planetesimals they accreted during formation \citep{PodolakEtAl-2022}. Here we elaborate on this problem, and propose a new potential solution. We show that chemical reactions between planetesimals dominated by organic-rich refractory materials and the hydrogen in gaseous atmospheres of protoplanets can form large amounts of methane 'ice'. Uranus and Neptune could thus be compatible with having accreted refractory-dominated planetesimals, while still remaining icy. Using random statistical computer models for a wide parameter space, we show that the resulting methane-rich internal composition could be a natural solution, giving a good match to the size, mass and moment of inertia of Uranus and Neptune, whereas rock-rich models appear to only work if a rocky interior is heavily mixed with hydrogen. Our model predicts a lower than solar hydrogen to helium ratio, which can be tested. We conclude that Uranus, Neptune and similar exoplanets could be methane-rich, and discuss why Jupiter and Saturn cannot.

\section{Introduction}\label{S:Intro}
The first models of Uranus and Neptune were constructed based on the premise that they were formed in the outer parts of a primitive proto-planetary nebula, where the elemental abundances were considered solar. Thus, assuming chemical equilibrium, early studies used calculations of the expected relative molecular abundances to gauge the potential composition of the planets that formed in this region \citep{ReynoldsSummers-1965,PodolakCameron-1974}. Ices, such as H$_2$O, CH$_4$, and NH$_3$ were collectively found to be about twice as abundant as rocky materials. Hereafter, we refer to rocks as those refractory minerals in which the ratio of C to silicate-forming elements such as Si, Fe, Mg or Ca is less than 0.1, whereas in refractory organics (CHONs) this ratio is larger than 10 \citep{FomenkovaEtAl-1992}. Refractory organics are a subset of organic compounds, however which sublimate at higher temperatures, typically in excess of 400 K \citep{NazariEtAl-2023}.
	
These early models typically considered a layered structure, with a rock core, an ice mantle, and an envelope rich in hydrogen and helium \citep{PodolakReynolds-1984,PodolakEtAl-1995}. Rocks were modeled mainly as silica (SiO$_2$), but sometimes mixed with other molecules (e.g., MgO, Fe, Ni) and water was considered the most abundant and therefore primary ice. The gas envelope was modeled mainly as a H+He solar mix with an admixture of H$_2$O, CH$_4$, and NH$_3$. Later developments highlighted the role of a more complex internal structure including material mixing and gradients \citep{HelledEtAl-2020,HelledFortney-2020}.
	
There is increasing evidence however that the constituent building blocks of the outer planets might be refractory-rich, rather than water-rich, challenging common convention with respect to Uranus and Neptune. We summarize these arguments below, indicating that refractory materials in the outer solar system could be 3-6 times more abundant than water.

\noindent\textbf{Intermediate-sized Kuiper Belt objects} (KBOs) indicate a trend of increasing bulk density with mass \citep{Brown-2012}. It was shown by means of several thermophysical evolution models (see appendix \ref{A:KBOs} for detailed references) that the trend may arise solely from differences in their internal porosity, due to thermal and mechanical processing. Assuming that KBOs have all formed in the same region and thus have a similar basic composition \citep{KenyonEtAl-2008}, these models have shown that a rock/water mass ratio of $\sim$3 fits the observed trend.\\

\noindent\textbf{Large KBOs} like Pluto are of such considerable size, that they contain very little residual porosity. Thus, its measured bulk density of 1854 kg m$^{-3}$ \citep{SternEtAl-2015} directly translates into a similarly rock-rich composition. Other large KBOs like Eris and Haumea often have even higher densities, which imply yet richer rocky compositions (although these are typically associated with further mantle-striping modifications via giant collisions \citep{LissauerStewart-1993,LeinhardtEtAl-2010,BarrSchwamb-2016}).\\

\noindent\textbf{Comets} are nowadays regarded as highly refractory-rich bodies. There is a growing body of evidence that strongly suggests comets have refractory to ice mass ratios of \emph{at least} 3, and potentially even twice that value \citep{RotundiEtAl-2015,FulleEtAl-2016,FulleEtAl-2017,FulleEtAl-2019,ChoukrounEtAl-2020}.\\

%\noindent\textbf{Molecular clouds} have a dust-to-water mass ratio close to 5 \citep{CambianicaEtAl-2020}, which at least partially entered into comets \citep{WoodenEtAl-2017}.\\

\noindent\textbf{Polluted white dwarfs} are astrophysical mass spectrometers. Their atmospheres probe the composition of accreted exo-Solar planetary debris, inferring the bulk chemical composition of the pollutants \citep{HollandsEtAl-2018,HarrisonEtAl-2018,HarrisonEtAl-2021}. The typical finding is of almost purely rocky debris \citep{JuraXu-2012,JuraYoung-2014,PietroGentileFusilloEtAl-2017}. It must be born in mind however that planetesimals can originate from different regions of the disk, devolatilize before their host star becomes a white dwarf \citep{MalamudPerets-2016,MalamudPerets-2017a,MalamudPerets-2017b}, and the progression of accretion can complicate the association between pollution and pollutant \citep{BrouwersEtAl-2022}. Despite these complications, relatively rare few individual cases point to polluting debris which probably originate in the outer parts of their systems and which contain a significant mass fraction of water \citep{FarihiEtAl-2013,RaddiEtAl-2015,XuEtAl-2017}. Yet even in these outlier cases the water fraction is 26\%-38\%, showing that a large rock/water mass ratio might be a ubiquitous feature of outer planetary disks.\\

These studies all suggest that the planetesimals accreted during the formation of Uranus and Neptune should contain \emph{at least} 2-3 as much mass in refractory material as in ices, leading to the question of why models of Uranus and Neptune seem to imply they are water-rich. This inconsistency, or composition tension problem, motivates our work, and there could only be two diametrically opposed solutions: (I) ice-poor building blocks can somehow lead to the formation of ice-rich planetary interiors, or (II) the internal structure models of Uranus and Neptune could be compatible with a large rock/ice mass ratio. We identify a novel path for enabling the former, transforming ice-poor building blocks into an ice giant by invoking chemistry. At the same time we discuss past literature that shows rock-rich interiors require a fine-tuned solution. Under ordinary assumptions, we demonstrate that the rock/ice mass ratio is very severely limited.

\section{Expected composition of outer solar system planetesimals}\label{S:CometComposition}
If small bodies in the outer solar system are indeed rich in refractory material, what do we expect this material to be composed of? We start by considering the detailed composition of solids that contributed to the accretion phase of the outer giant planets. However, as we have very little direct information about the mineralogy of nearby Kuiper belt solids, we assume instead accretion of solids with comet-like composition, since comets similarly originated in the outer solar system. This assumption was also made in some prior studies of Uranus and Neptune \citep{HubickyjEtAl-2005,IaroslavitzPodolak-2007}.

Present-day comets typically contain tiny amounts of CO and other hypervolatiles, relatively little water, and are \emph{primarily} composed of refractory materials \citep{BockeleeMorvanBiver-2017,FulleEtAl-2017,FulleEtAl-2019,ChoukrounEtAl-2020}. Observations of various proto-planetary discs indirectly imply that when sufficiently far out in the disc, carbon is depleted from the gas phase and instead enters into solid grains locked inside planetesimals \citep{McClure-2019,TaboneEtAl-2023}. Below we discuss \emph{direct} observations of cometary dust from our own solar system, pointing towards a composition rich in carbon-bearing organic refractories. We show that between 1/3-1/2 of the mass of their refractory material is contained in organic rather than silicate material:\\

\noindent\textbf{Comet P/Halley} was explored by \textit{Giotto}, a flyby mission in 1986 \citep{KellerEtAl-1986}, showing that cometary dust is extremely rich in organics relative to typical chondrites, and more similar to dust in dense molecular clouds. High-velocity collisions of dust particles with metal targets on the spacecraft provided a source of ions for mass-spectrometric analysis \citep{KisselEtAl-1986,LawlerBrownlee-1992}. Particles containing higher fractions of organics were relatively more abundant closer to the nucleus \citep{Fomenkova-1999}, since organic components are more easily dissociated in the coma \citep{FomenkovaEtAl-1994,WoodenEtAl-2007}, providing an extended source for various gaseous species \citep{CottinFray-2008}. Studies also discuss the detailed distribution of carbon among different compounds \citep{Fomenkova-1999}. Overall, the mass ratio of rocks to organics in dust is between 2 and 1 \citep{FomenkovaEtAl-1992,Fomenkova-1999}; i.e, carbon-rich organics constitute as \textbf{33-50\%} of the dust mass.\\

\noindent\textbf{Comet 67P/Churyumov-Gerasimenko} was explored by ESA's \textit{Rosetta} mission, which was the first to successfully rendezvous with a comet nucleus, follow it in its orbit and dispatch a lander to the surface. The first spectrospcopic detection of organics on the surface of a comet was made by the Rosetta VIRTIS near-IR spectrometer instrument. Carboxylic acid was suggested as a primary organic material on the nucleus \citep{CapaccioniEtAl-2015,QuiricoEtAl-2016}. The Rosetta COSIMA instrument is a secondary ion mass spectrometer equipped with a dust collector, a primary ion gun, and an optical microscope for determining the position and characteristics of the dust particles \citep{LeRoyEtAl-2012}. Many organic compounds were identified \citep{LeRoyEtAl-2015}, and approximately \textbf{45\%} of the refractories in comet 67P/C-G could be in organics \citep{BardynEtAl-2017}.\\

\noindent\textbf{Comet C/2013 US$_{10}$ (Catalina)}, a dynamically new Oort cloud comet with a close-Earth approach, was determined to have a \textbf{47\%} amorphous carbon mass fraction \citep{WoodwardEtAl-2021}.\\

\noindent\textbf{Ultra-carbonaceous Micrometeorites} (UCAMMs) represent a small fraction of interplanetary dust particles (IDPs) reaching the Earth, and are exceedingly carbon rich with respect to other types of IDPs \citep{ThomasEtAl-1993,ThomasEtAl-1994}, as well as meteoritic samples \citep{DartoisEtAl-2018}. They are recovered in the antarctic continent in a relatively pristine state \citep{DupratEtAl-2010}. The bulk fraction of organic refractory material in UCAMMs ranges between \textbf{22-58\%} \citep{DobricaEtAl-2012}. A recent review \citep{WoodenEtAl-2017} pieces together a line of arguments that advocate in favor of a cometary origin for UCAMMs: (a) their crystalline and amorphous silicates are similar to those in comet dust \citep{Wooden-2008}; (b) high S and low Si abundances in glass with embedded metal and sulfides promote a cometary origin \citep{BradleyIshii-2008}; (c) their fluffy textures appear minimally altered by heating during relatively low-velocity atmospheric entry usually associated with cometary origin \citep{DupratEtAl-2010}; (d) they have D/H ratio enrichment expected in cometary dust \citep{DupratEtAl-2010}, including that analyzed by Rosetta \citep{PaquetteEtAl-2021}; (e) they are globally unequilibrated, in a similar manner as other cometary materials (e.g from \textit{Stardust} samples \citep{BrownleeEtAl-2007}). All this taken together indicates a cometary-like nature for UCAMMs. We also note briefly that many similar arguments also place the origin of relatively carbon-rich chondritic anhydrous IDPs in comets \citep{EngrandEtAl-2023}.\\

\noindent\textbf{\textit{Spitzer} remote observations} of cometary dust revealed a wide range of amorphous carbon mass fractions spanning 10-90 \%, with a mean value of \textbf{54\%}, based on a large set of a few dozen comets \citep{HarkerEtAl-2023}. This work supports a new paradigm in which most comets have a carbonaceous content of $\sim$50\%.\\

In Appendix \ref{A:Stardust} we provide a note about why \textit{Stardust}, the only comet sample return mission to date, could not reliably provide an assessment of the bulk organic fraction. However, enough evidence has accumulated as portrayed above, in order to establish carbon-rich organics as major constituents of the solids that accreted to form the outer planets. We will later discuss why such solids can be transformed into copious quantities of methane in Uranus' and Neptune's bulk interiors, during their growth.

\section{Expected composition of Uranus and Neptune}\label{S:UranusNeptuneComposition}
As noted in the introduction, classical models of Uranus and Neptune assumed they were ice-rich, based on the expected composition of a solar-composition gas in equilibrium at low temperature.  Models constructed based on this assumption were found to be consistent with the observed physical properties of these planets. If, as we argued in the previous section, the planetesimal building blocks have a high refractory content, why do Uranus and Neptune appear to be water rich? In order to investigate this composition tension problem, we explore below the various possibilities for their composition using a statistical approach.

We have previously developed \citep{PodolakEtAl-2022} and here further modified a statistical computer code that creates random, monotonic density distributions, which fit a given mass, radius, and moment of inertia. For Uranus and Neptune the gravitational moments J$_2$ and J$_4$ have also been measured \citep{Jacobson-2009,Jacobson-2014}. However, in order to compute these moments for a given density profile, the rotation period of the body must be known, and for both Uranus and Neptune there is some ambiguity in this matter \citep{HelledEtAl-2011}. In addition, the computation of these moments is extremely time-consuming, so we have chosen, instead, to fit the moment of inertia (MOI) only. While less accurate than using J$_2$ and J$_4$, the difference in the derived density distribution is not significant for our purpose \citep{PodolakEtAl-2022}.

By integrating the equation of hydrostatic equilibrium we can find the pressure for each density. \cite{PodolakEtAl-2022} describe several algorithms for associating a temperature and composition with each pressure-density point. Here we use a modified version of the 'inward' algorithm. Briefly, we assume that at any point the material is composed of a mixture of two adjacent components from the following set of six materials, ordered according to intrinsic density: Fe, SiO$_2$, CO, H$_2$O, CH$_4$, and a solar mix hydrogen/helium gas (hereafter called \emph{envelope}). Thus SiO$_2$ can mix with either Fe or CO (or H$_2$O for models which do not include carbon compounds), but not with anything else. In addition, the mass fraction of the denser component cannot decrease going inwards. \cite{PodolakEtAl-2022} did not consider CO and CH$_4$, and we now allow for these species as well due to the importance of carbon in outer solar system composition. Starting with a given surface temperature and pressure, we compute the mass fractions of envelope and CH$_4$ (or H$_2$O for models which do not include carbon compounds). Continuing inwards, we try to minimize the mass fraction of the denser material while keeping the temperature constant. If there is no mix that matches under those conditions, we increase the temperature while keeping the same composition. The 'inward' algorithm tries to maximize the rock to water ratio while minimizing the central temperature. Nevertheless, the central temperatures that are generated can sometimes be quite high, and we discard those models where it exceeds $2.5 \times 10^4$ K, which we consider to be unlikely based on models of heat transport in these planets \citep{PodolakEtAl-2019,VazanHelled-2020}. 

For computing random models with Fe (iron), SiO$_2$ (silica) and H$_2$O (water) we use the quotidian equation of state \citep{MoreEtAl-1988,VazanEtAl-2013}. For H+He (envelope) material we use SCvH equation of state \citep{SaumonEtAl-1995} as modified in \cite{ChabrierEtAl-2019}. For CO (carbon-monoxide) we use the quotidian equation of state \citep{PodolakEtAl-2023}. For CH$_4$ (methane) we use the well-known SESAME equation of state \citep{LyonJohnson-1992}. For mixing of different materials we use the additive volume law.

By generating hundreds of thousands of models, we can explore the space of allowable internal compositions and structures. While we cannot say which internal composition is more likely, we can in fact place meaningful limits on the composition.

\subsection{Water-rich Models}\label{SS:Water-rich}
\cite{PodolakEtAl-2022} investigated Uranus models with no carbon and with a MOI factor of 0.23. They found that, for central temperatures below $2.5 \times 10^4$\,K there were no models with a rock/water mass ratio of 2 or more, although some models came quite close. Thus one model, with a rock/water mass ratio of 1.90 had 6.63\,$M_{\oplus}$ of rock, 3.49\,$M_{\oplus}$ of water, and 4.41\,$M_{\oplus}$ of envelope. The central temperature was $T_{\rm c}=1.77 \times 10^4$\,K. A second model, with a rock/water mass ratio of 1.85 had 6\,$M_{\oplus}$ of rock, 3.25\,$M_{\oplus}$ of water, and 5.29\,$M_{\oplus}$ of envelope. For this model $T_{\rm c}=1.14 \times 10^4$\,K. A feature of many of these relatively rock-rich models is that they reach much higher central temperatures compared to water-rich models.
%is that immediately surrounding the Fe+rock core is a thin layer rich in water. Above this is a region with roughly equal mass fractions of water and envelope. Near the outer edge of the planet is a region rich in envelope material. Thus there are essentially four layers with composition gradients, and typically very high $T_{\rm c}$.

Newly generated results in this study confirm previous results \citep{PodolakEtAl-2022} (and are therefore not shown here explicitly), indicating that the rock/water mass ratio of both Uranus and Neptune is severely restricted when typical materials are considered, which include Fe, SiO$_2$, H$_2$O, and H+He envelope gas. The exact upper limit value depends sensitively on which moment of inertia factor was assigned to Uranus and Neptune. Whereas classical models considered a MOI factor of 0.23 for both planets \citep{dePaterLissauer-2015}, the most recent studies \citep{NeuenschwanderHelled-2022} suggest that 0.22 is probably a more appropriate value for Uranus and 0.24 for Neptune. We now considered these newer values as lower and upper limits, respectively. Using these new values, the rock/water mass ratio obtained statistically is likewise never more than 2, and actually rarely exceeds 1. In contrast, the water/rock mass ratio has no upper bound, and there are a multitude of water-rich models with far lower central temperatures. These results are consistent with past studies, which almost never found large rock/ice mass ratios.

\subsection{Rock-rich Models}\label{SS:Rock-rich}
It should be noted that, perhaps, more rock-rich interiors are possible using future algorithms which will consider other material mixing options. Whereas classic layered models often did not consider mixing at all (especially in the rock component), we have considered mixing of multiple adjacent materials, while some studies \citep{HelledEtAl-2011,VazanHelled-2020} also considered mixing of SiO$_2$ and H+He, or triple mixing of SiO$_2$, H$_2$O and H+He \citep{VazanEtAl-2020}. Despite these various options, neither \cite{VazanHelled-2020} nor any other study has ever reached a rock/water mass ratio larger than 2, with one exception. \cite{HelledEtAl-2011} managed to accomplish that by suggesting a mixture of pure SiO$_2$ with a very large fraction (15\%) of H+He. While a rock-rich interior is therefore technically possible, recent discussion regarding the question of miscibility \citep{NettelmannEtAl-2016,MiguelVazan-2023} supports the adjacent mixing of SiO$_2$ and H$_2$O, while mixing of SiO$_2$ and hydrogen (And also Fe and hydrogen, see \cite{WahlEtAl-2013,Komabayashi-2021}) remains empirically under-explored and probably requires high temperatures. While our conservative choice of adjacent mixing only is a general limitation, other modes of mixing are perhaps more uncertain and require fine-tuned conditions to obtain a 'desirable' outcome that leads to rock-rich interiors.

\subsection{Models with Carbon Monoxide}\label{SS:CO_Models}
When one considers ices in the outer solar system, one typically neglects ices which are not water, since water is by far the most abundant ice in present day comets \citep{BockeleeMorvanBiver-2017}. However, CO is the second-most abundant volatile in molecular clouds \citep{TanakaEtAl-1994,ChiarEtAl-1995}, and in the early solar system, we expect to have a large fraction of condensed carbon monoxide ice, even with respect to water \citep[][and the references therein]{LisseEtAl-2022}. At a later stage, CO ice in the region of Uranus and Neptune is expected to sublimate \citep{LisseEtAl-2022}, and eventually disperse along with the gaseous proto-planetary disc \citep{ErcolanoPascucci-2017}. A possible way around this, is if CO ice survives inside comets that are placed into distant Oort cloud orbits early enough in order to avoid complete loss of CO ice \citep{Davidsson-2021}. C/2016 R2 might be a prime example of a rare, extremely hyper-volatile rich, yet water and dust poor comet, belonging to this category \citep{McKayEtAl-2019}. Such comets provide an intriguing indication that the solar system may have indeed began with abundant CO, as hypothesized above.

Assuming that CO was indeed prevalent when Uranus and Neptune formed, and also ignoring chemistry and additional concerns, which are addressed later in the paper, we explore the space of possible compositions for Uranus and Neptune given two ices - water and carbon monoxide. If the CO can replace the H$_2$O in these models, it should raise the rock/water mass ratio above 2. For comparison purposes, we use an identical numerical setup as in our previous study of randomly generated models \citep{PodolakEtAl-2022}, only now instead of 4 materials (iron, silica, water and H+He gas), we investigate 5 materials, with carbon monoxide as an additional ice. We consider both Uranus and Neptune and account for MOI factors in the range 0.22-0.24, as previously outlined.

\begin{figure}[h!]

	\begin{tabular}[b]{c}
		\includegraphics[scale=0.53]{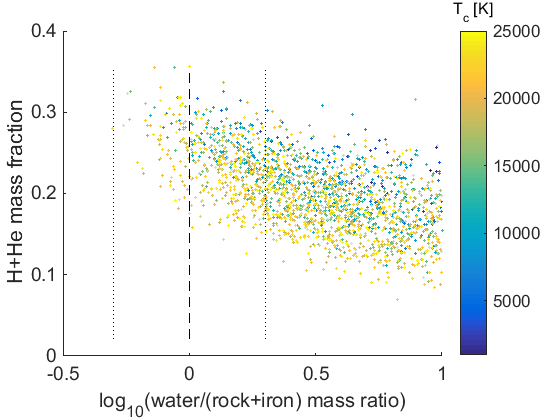}\label{fig:env_watertorock_CO_Ura_22.png}\\
		\small (a) Uranus MOI factor = 0.22
    \end{tabular}
    \begin{tabular}[b]{c}
 	    \includegraphics[scale=0.53]{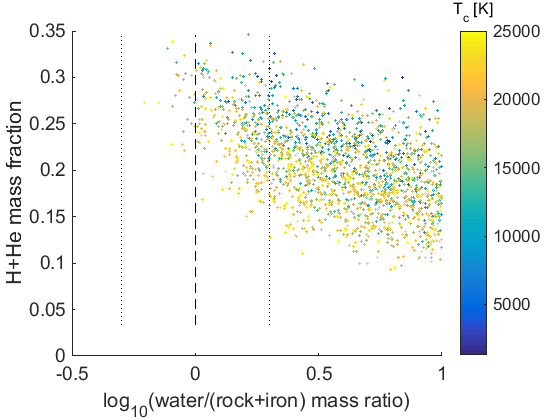}\label{fig:env_watertorock_CO_Ura_23.png}\\
 	    \small (b) Uranus MOI factor = 0.23
    \end{tabular}
    \begin{tabular}[b]{c}
		\includegraphics[scale=0.53]{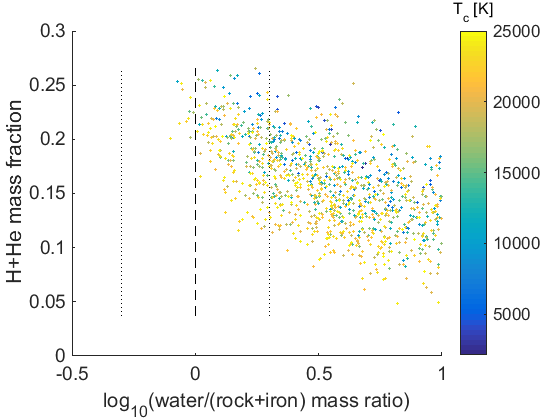}\\
		\small (c) Neptune MOI factor = 0.23
	\end{tabular}
	\begin{tabular}[b]{c}
        \includegraphics[scale=0.53]{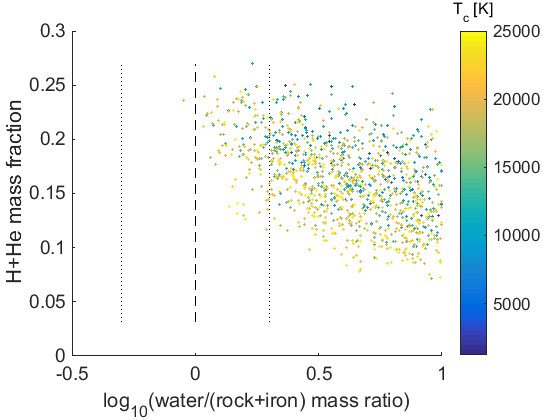}\\
		\small (d) Neptune MOI factor = 0.24
	\end{tabular}
	\caption{Mass fraction of envelope vs. water/rock mass ratio (logarithmic) in models of Uranus and Neptune that include CO as an additional ice to water. Each point denotes a random interior model, color coded by its central temperature. The dashed vertical line indicates equal fractions of water and rock. The right (left) dotted vertical line indicates a water/rock mass ratio of 2 (0.5).}
	\label{fig:COenv_vs_watertorock}
\end{figure}

Using the inward algorithm to match 100000 randomly generated density profiles, we show the interior models of Uranus and Neptune with CO in Figure \ref{fig:COenv_vs_watertorock}. Each random model is depicted by a single pixel, whose color corresponds to the central temperature. The mass fraction of envelope in the planet is shown as a function of the water/rock mass ratio. We note that by rock we refer to the high-Z components SiO$_2$+Fe, and by ice we refer to H$_2$O+CO (distinguishing ice/rock from water/rock). Since here the water/rock mass ratio extends to very large values, we obtain a better view when the x-axis is in logarithmic scale and truncated at a water-dominated mass ratio of 10.

\begin{figure}[h!]	
	\begin{tabular}[b]{c}
		\includegraphics[scale=0.53]{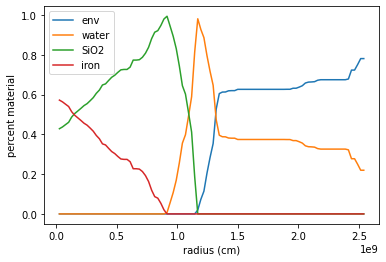}\label{fig:ExampleWithoutCO}\\
		\small (a) Before adding CO
	\end{tabular}
	\begin{tabular}[b]{c}
		\includegraphics[scale=0.53]{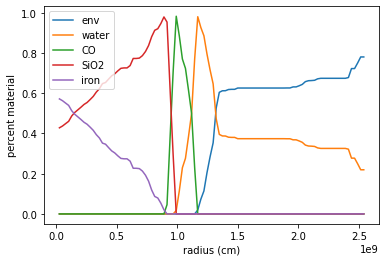}\label{fig:ExampleWithCO}\\ 
		\small (b) After adding CO
	\end{tabular}
	
	\caption{An example of a random rock-rich model of Uranus, and the internal delineation of its composition (normalized), before and after adding CO ice as a fifth material. This model has a MOI factor of 0.22, central temperature of 12000 K and water/rock mass ratio of 0.7.}
	\label{fig:CO_internal_composition}
\end{figure}

The dashed vertical line denotes equal proportions of water and rock. The right (left) dotted vertical line indicates a water/rock mass ratio of 2 (0.5). Models with zero rock ($\infty$ on the x-axis) are not shown. As can be seen, much of the model parameter space is in the region of large water/rock mass ratios. As was the case for the models without CO, there are no models with a water/rock mass ratio less than 0.5, that is, the rock/water mass ratio is never more than 2, and rarely exceeds 1. The reason for this is that, in addition to replacing some H$_2$O, the CO predominantly replaces SiO$_2$, as can be seen in Figure \ref{fig:CO_internal_composition}.

CO has a density between that of water and SiO$_2$. Therefore the compositional layers transition smoothly from water to CO and then SiO$_2$. If the same density profile is fit first without CO ice and then including it, CO can replace either water or SiO$_2$. In the models CO tends to replace more of the SiO$_2$, than the water. Thus, in Figure \ref{fig:CO_internal_composition}, we see the mass fraction of each component as a function of radius for a particular Uranus model, originating from the top left of Figure \ref{fig:COenv_vs_watertorock}, panel (a). Here the MOI is 0.22, the central temperature reaches 12000 K and the water/rock mass ratio is 0.7. Comparing the model with CO (right panel) to that without (left panel) we see that the CO tends to replace the SiO$_2$ in preference to the water. Thus, simply adding CO to models is not a good solution for the composition tension problem of Uranus and Neptune.% Additionally, after having explored so far two options for the the ice composition (either with or without CO), in both of these cases it is easy to find many water-rich models with low central temperatures (several 10$^3$ K), however in the rock-rich models the central temperatures are often larger than $\sim$few 10$^4$ K. Water-rich models are thus much more consistent with theoretical expectation from the initial heat sources and subsequent thermal evolution \citep{PodolakEtAl-2019,PodolakEtAl-2022}.

%We also find that lowering the MOI leads to models that are more rock-rich. This is consistent with the fact that a greater central concentration of mass (lower MOI) implies a larger rock core. The Neptune models tend to be more water-rich. Presumably this is related to the fact that Neptune's higher density tends to yield models with a lower mass fraction of envelope. The rock-rich models (left of the dashed line) share a similar property of having an envelope mass fraction close to that maximum value. Thus, given the same MOI factor (here 0.23, panels b and c), Uranus with its larger low-Z atmosphere compared to Neptune can compensate by generating random density profiles that are centrally denser, in order to reach the same MOI value. This intuitively explains why its limit rock/water mass ratio is higher.

\subsection{Models with Methane}\label{SS:CH4_models}
We also consider a third ice, methane, as motivated by the following arguments. The CH$_4$ abundance in the upper atmosphere of Uranus and Neptune is observed to be far greater than solar \citep{AtreyaEtAl-2020}. This should not be regarded as indicative of the \emph{bulk} abundance of methane, since these measurements only probe the atmosphere at the $\sim$1 bar level. Nevertheless, early studies did in fact consider models where CH$_4$ was mixed into the entire envelope with far greater abundances than solar \citep{Podolak-1976,PodolakReynolds-1981}, or with an icy layer that consisted primarily of H$_2$O and CH$_4$ \citep{HubbardMacfarlane-1980}. This was motivated by the idea that nebular carbon should condense to form CH$_4$ and accrete onto Uranus and Neptune in large amounts. At around the same time, however, the view was shifting away from condensation of solid CH$_4$ as a result of kinetic inhibition \citep{LewisPrinn-1980}. Instead, at sufficiently low temperatures, the expectation was that the large majority of carbon would condense directly to CO \citep{LewisPrinn-1980}.
	
Later it was briefly suggested that, instead, CO might be chemically transformed in the atmosphere of the gas giants and convert to CH$_4$ \citep{PodolakReynolds-1984} when reacting with hydrogen, altering their bulk composition. However, this hypothesis did not develop further since. Here we extend the idea that CH$_4$ could be a key volatile in the outer planets. As already outlined in Section \ref{S:CometComposition}, while CO is almost negligible and therefore has a very small potential contribution for generating methane in typical contemporary planetesimals, refractory organics contribute to a very significant portion of their mass. These refractory organics may also chemically transform into copious amounts methane when they accrete in vast quantities onto the planet's atmosphere, and are capable of altering its bulk composition.

\begin{figure}[h!]
	\begin{tabular}[b]{c}
		\includegraphics[scale=0.53]{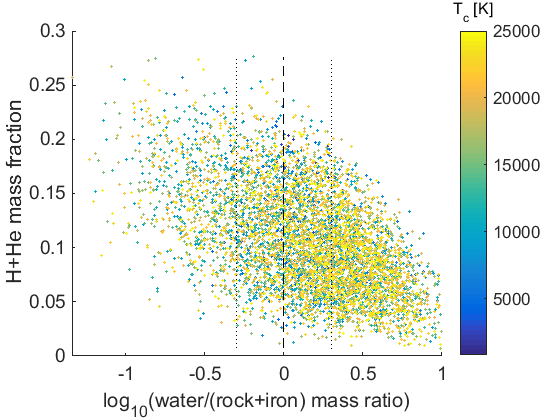}\label{fig:env_watertorock_CH4_Ura_22}\\
		\small (a) Uranus, MOI factor = 0.22
	\end{tabular}
	\begin{tabular}[b]{c}
		\includegraphics[scale=0.53]{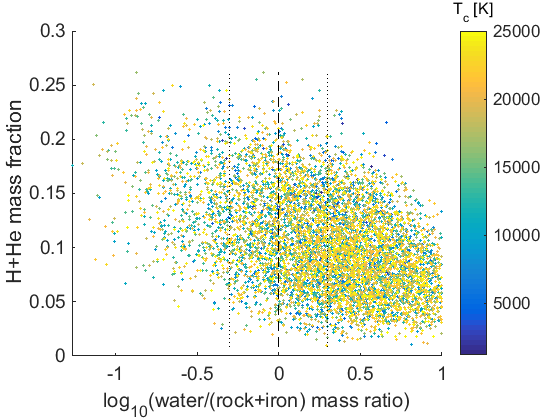}\label{fig:env_watertorock_CH4_Ura_23}\\ 
		\small (b) Uranus, MOI factor= 0.23
	\end{tabular}
	\begin{tabular}[b]{c}
		\includegraphics[scale=0.53]{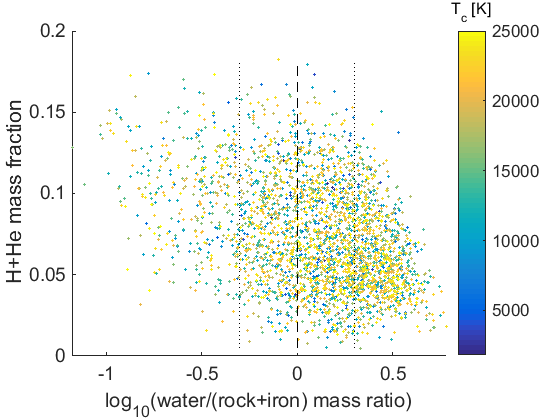}\label{fig:env_watertorock_CH4_Nep_23}\\
		\small (c) Neptune, MOI factor = 0.23
	\end{tabular}
	\begin{tabular}[b]{c}
		\includegraphics[scale=0.53]{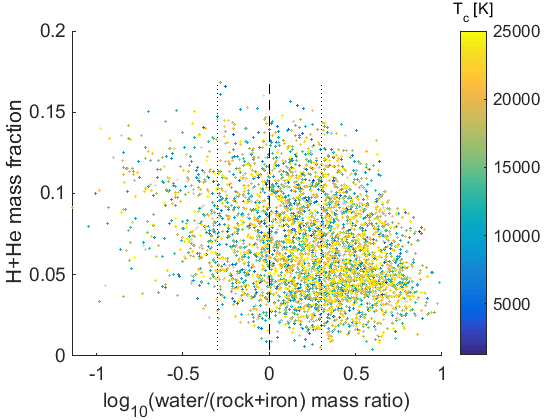}\label{fig:env_watertorock_CH4_Nep_24}\\
		\small (d) Neptune, MOI factor = 0.24
	\end{tabular}
	
	\caption{Same as Figure \ref{fig:COenv_vs_watertorock}, but for models with CH$_4$ as an additional ice to water, instead of CO.}
	\label{fig:CH4env_vs_watertorock}
\end{figure}

We use the same setup as in the previous section, but now replace carbon monoxide with methane. When we incorporate CH$_4$ as our second icy material, the results are very different. Randomly generated interior models of Uranus and Neptune with CH$_4$ are shown in Figure \ref{fig:CH4env_vs_watertorock}. In these models the water/rock mass ratio now spans the range 0.1-10. As noted above, CO as an additional ice tends to replace SiO$_2$. CH$_4$, on the other hand, tends to replace water, so the rock/water mass ratio can be significantly larger.

\begin{figure}[h!] 
	%	\setcaptiontype{figure}
	\centering
	\begin{tabular}[b]{c}
		\includegraphics[scale=0.53]{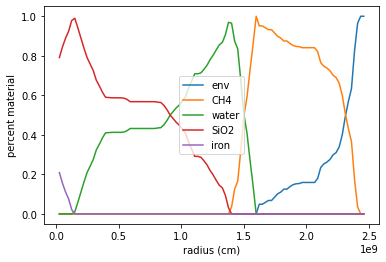}\label{fig:ExampleWithCH4_comp}\\
		\small (a) Compositional delineation
	\end{tabular}\\
	\begin{tabular}[b]{c}
		\includegraphics[scale=0.53]{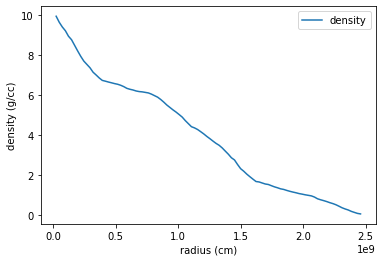}\label{fig:ExampleWithCH4_density}\\ 
		\small (b) Density profile
	\end{tabular}\\
	\begin{tabular}[b]{c}
		\includegraphics[scale=0.53]{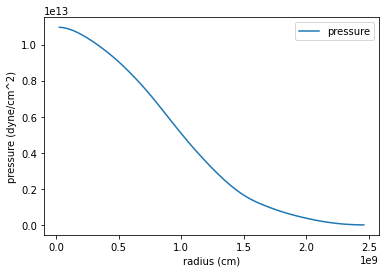}\label{fig:ExampleWithCH4_pressure}\\ 
		\small (b) Pressure profile
	\end{tabular}
	
	\caption{An example of a typical random water-rich model of Neptune}
	\label{fig:CH4_internal_composition}  % seed 105.037177876
	\end{figure}

Is methane stable to pressure-induced dissociation in Uranus' and Neptune's interior? Methane has a density between that of envelope H+He gas and water. Therefore the compositional layers transition smoothly from envelope to methane and then water. Figure \ref{fig:CH4_internal_composition} shows the mass fractions of the different materials, the density, and the pressure as a function of radius for one of Neptune's random models. In this case the MOI factor is 0.24, the central temperature reaches 13700 K and the water/rock mass ratio is 2.

In panel (a) we see the transition from methane-dominated to water-dominated composition, at a distance of about 15000 km from the planet's center ($\sim$60\% of the total radius). The corresponding pressure at this point is shown in panel (c) to be just under 200 GPa, which is less than the 300 GPa required for complete methane dissociation \citep{BenedettiEtAl-1999}. A certain fraction of methane may reside in the pressure range 100-300 GPa, which might be enough to partially dissociate some of it into other hydrocarbons \citep{AncilottoEtAl-1997,BenedettiEtAl-1999,GaoEtAl-2010,ShermanEtAl-2012}, but should not result in complete dissociation into pure carbon and hydrogen. In addition, since methane and water can be mixed in the transition layer, it should be noted that pressures up to 150 GPa can also give rise to a high-pressure form of methane clathrate hydrate \citep{SchaackEtAl-2019}. We have examined a number of random models in detail, and the results are similar. Typically, only the deepest submersed methane is just marginally able to dissociate due to high pressure. We thus consider our assumption of using the methane equation of state as judicious, and we expect little difference in density at these pressures before and after dissociation.

\begin{figure}[h!]
	%	\setcaptiontype{figure}
	\begin{tabular}[b]{c}
		\includegraphics[scale=0.53]{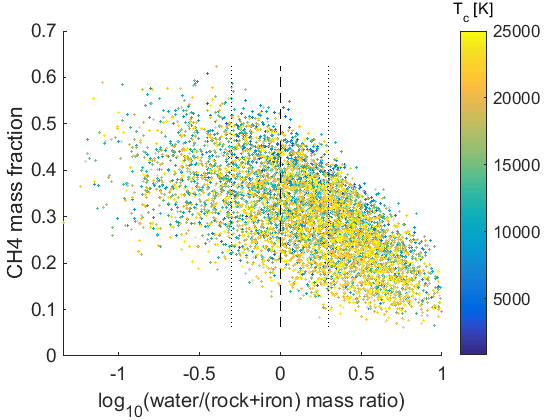}\label{fig:ch4_watertorock_CH4_Ura_22}\\
		\small (a) Uranus, MOI factor = 0.22
	\end{tabular}
	\begin{tabular}[b]{c}
		\includegraphics[scale=0.53]{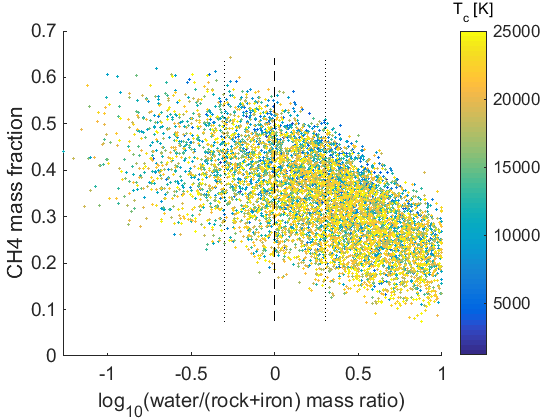}\label{fig:ch4_watertorock_CH4_Ura_23}\\ 
		\small (b) Uranus, MOI factor = 0.23
	\end{tabular}
	\begin{tabular}[b]{c}
		\includegraphics[scale=0.53]{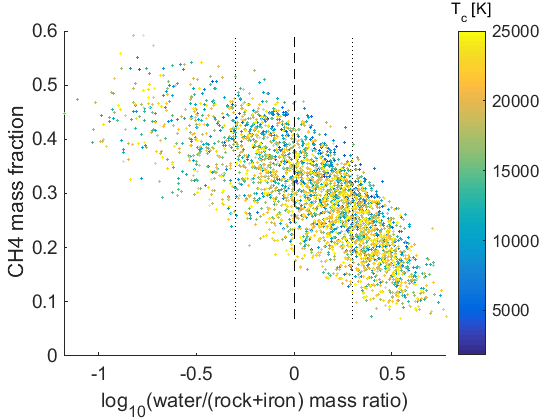}\label{fig:ch4_watertorock_CH4_Nep_23}\\
		\small (c) Neptune, MOI factor = 0.23
	\end{tabular}
	\begin{tabular}[b]{c}
		\includegraphics[scale=0.53]{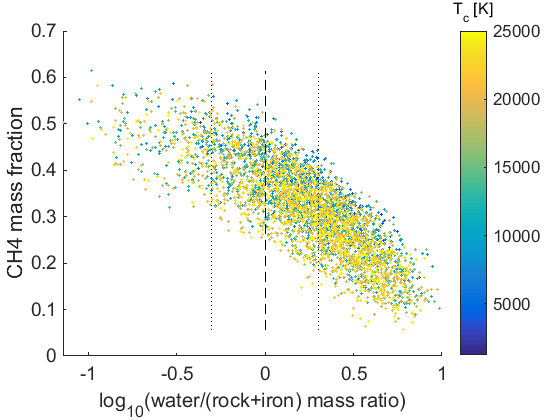}\label{fig:ch4_watertorock_CH4_Nep_24}\\
		\small (d) Neptune, MOI factor = 0.24
	\end{tabular}
	
	\caption{Same as Figure \ref{fig:CH4env_vs_watertorock}, but showing the CH$_4$ mass fraction instead of envelope mass fraction.}
	\label{fig:ch4_vs_watertorock}
	\end{figure}

Finally, in Figure \ref{fig:ch4_vs_watertorock}, we show the methane mass fraction as a function of the water/rock mass ratio. The methane mass fraction probed by the space of possible solutions is at least 10\%. In models that also have a large rock/water mass fraction (left of the vertical dashed line) it is necessarily larger, upwards of 20\%. Since the envelope mass fraction in these models also has large values (Figure \ref{fig:CH4env_vs_watertorock}), the combination of large envelope and even larger CH$_4$ mass fractions means that the other materials - H$_2$O, SiO$_2$, and Fe, are collectively only a relatively small fraction of the planet's mass. Thus, in models that have a large rock/water mass fraction, specifically, the methane fraction might even be larger than that of H$_2$O, SiO$_2$, and Fe combined. This is a difficulty because CH$_4$ is certainly not that prevalent in the contemporary solar system, and is normally a very small fraction compared to water \citep{BockeleeMorvanBiver-2017}. Thus, for methane to participate in solving the composition tension problem, a large methane fraction must be justified. 

We note that since in our current model we only mix two adjacent materials, it might be possible that other mixing choices would result in less methane. For instance, triple mixing of water, methane and H+He in the outer atmosphere might lower the bulk methane fraction. On the other hand, it seems reasonable to expect that if our current water + methane region and methane + envelope region were to be mixed into one homogeneous region, the overall material masses would not change dramatically. It remains a task for future studies to quantitatively explore the triple mixing possibility, but the methane fraction will likely remain too large to reconcile with its contemporary abundance in the solar system, either way.

We suggest instead that organic-rich refractories are sufficiently abundant in outer solar system planetesimals to drive chemical reactions in the atmospheres of Uranus and Neptune, which, during the phase of planet growth, may produce the required copious amounts of methane. These processes are discussed next.

\section{Growth of the Outer Planets and Accretion of Solids}\label{S:Growth}
We consider a growing planet in the outer solar system, which starts accreting mass onto an embryo-sized core (the incipient core mass in the range 0.05-0.5 M$_\oplus$ \citep{BrouwersEtAl-2018,BodenheimerEtAl-2018}). Classical models emphasized subsequent growth due to accretion of large planetesimals. These planetesimals could fall directly onto the core \citep{PodolakEtAl-1988} however they could also dissolve in the atmosphere due to thermally \citep{IaroslavitzPodolak-2007,BrouwersEtAl-2018} or mechanically \citep{MordasiniEtAl-2015} driven ablation. As core mass accumulates, additional atmosphere is collected \citep{PollackEtAl-96}. Energy release is added to the luminosity of the planet and thus internal temperatures increase during growth \citep{PollackEtAl-96}.

Planetesimals can be large enough so that thermal ablation alone will not erode a significant fraction of their volume \citep{VallettaHelled-2019}. However, ram-pressure can fragment large planetesimals which are insufficiently strong. In that case they break up and become pancaked, decelerating much more effectively while passing through the atmosphere, and quickly lose their kinetic energy before reaching the core. If they are sufficiently large, exceeding about 10-100 km, this outcome might be avoided due to high self-gravity, and direct impacts onto the core are possible \citep{MordasiniEtAl-2015,VallettaHelled-2019}.

A second accretion process that was discussed in the literature involved pebble accretion \citep{LambrechtsEtAl-2014,BitschEtAl-2015,Ormel-2017,OrmelEtAl-2021}. Pebbles are small mm-dm sized objects, and as such they are more susceptible to gas drag. The planet's effective cross section for accretion is much larger. The ablation of the pebbles in the atmosphere is also greatly enhanced \citep{BrouwersEtAl-2018,OrmelEtAl-2021}. In reality, the accretion of Uranus and Neptune might have involved both pebbles and planetesimals, and thus we examine both options.

Pebbles are much more easily affected by thermal heating due to their small size and the increase in ambient temperature as they plunge deeper into the atmosphere. They are also affected by frictional heating, when a fraction of the kinetic energy lost through friction is converted into heat. Frictional heating is more dominant in the cold outer regions of the atmosphere, particularly for fast impacting pebbles. As a result, they quickly and fully evaporate and dissolve into the atmosphere \citep{BrouwersEtAl-2018}. Despite the fact that ablation is more efficient for pebbles than for planetesimals, both types of objects are viable sources for dispersing high-Z materials into the atmosphere \citep{BrouwersEtAl-2018}, and we consider them by accounting for a range of object sizes: 10 m, 100 m, 1 km, 10 km and 100 km. Solid objects below 10 m are eroded so effectively that we need not consider them explicitly. The exact details of ablation of both small and large planetesimals depend sensitively on material properties and especially the atmospheric density profile, which changes during the planet's growth.

We can get an idea of the relevant density profile, as well as the temperature-pressure (T-P) profiles, using a newly updated code \citep{StevensonEtAl-2022} implemented here for the first time for a growing Uranus. This code solves the equations of mass conservation, hydrostatic equilibrium, energy conservation, and energy transport assuming spherical symmetry. It includes the effects of mass and energy deposition by accreting solids and gases, and equations of state of water, SiO$_2$ and iron. It also includes opacity due to dust grains, the ablation and breakup of (here only 100 km mono-sized) solid planetesimals in the gaseous envelope, treatment of compositional gradients and heating by the host star which drives mass loss. Using this code, we obtain snapshots of the density/temperature/pressure profiles of the atmosphere, at three different times after the start of the evolution: 1.35, 2.06 and 2.65 Myr (see Figure \ref{fig:BodenheimerCode}).

We then use these atmospheric density profiles as input to an ablation code \citep{PodolakEtAl-1988}, which includes the detailed effects of frictional gas drag, heating by ambient thermal radiation, and the ensuing evaporation (sublimation) erosion derived from the energy input. Following the trajectories of accreted planetesimals of various sizes, we can determine the characteristic temperature-pressure conditions during ablation. For the parameters we selected, we find that small sub-km planetesimals are strongly slowed by atmospheric drag and ablate as a consequence. The larger planetesimals are correspondingly less affected by drag and are therefore able to maintain the speed necessary to break up deeper in the atmosphere, due to ram pressure. As already mentioned, these results depend on the assumed material properties. Since the code is unable to account for the intricacies of planetesimal intra-porosity or material mixing, we employ a similar approach to other studies \citep{IaroslavitzPodolak-2007,BrouwersEtAl-2018,BrouwersEtAl-2020,OrmelEtAl-2021} and simplify matters by considering non-porous planetesimals with a uniform density of 2000 kg m$^{-3}$, and the material thermodynamic/cohesion properties of water, which might represent a mean value for a mixture of silicates, organics, water and other high-volatility ices. Given these assumptions, the temperature-pressure conditions during ablation are provided in Table \ref{tab:ablationTP}. Note that ablation is a gradual process and therefore it occurs for a range of T-P conditions. To simplify matters, for sub-km planetesimals we listed a single T-P value that matches the moment during which half of the planetesimal mass has eroded. For larger planetesimals, the listed value matches the moment of ram-pressure breakup when most of the mass is instantantly delivered to the atmosphere.

\begin{table*}[h!]
	\caption{Temperature and pressure conditions during ablation of planetesimals, when, for sub-km planetesimals, half of the planetesimal mass has eroded; and for the larger planetesimals ram-pressure breakup occurs.}
	\centering
	\smallskip
	%\begin{minipage}{14.2cm}
	\begin{tabular}{|l|l|l|l|l|l|}
		\hline
		\textbf{Time~\textbackslash~Size} & 10 m & 100 m & 1 km & 10 km & 100 km \\ \hline
		
		1.35 Myr & 300 K, 90 Pa & 500 K, 1 kPa & 500 K, 2 kPa & 500 K, 2 kPa & 2300 K, 700 kPa \\
		2.06 Myr & 290 K, 40 Pa & 470 K, 500 Pa & 950 K, 20 kPa & 950 K, 20 kPa & 1900 K, 400 kPa \\
		2.65 Myr & 230 K, 20 Pa & 410 K, 400 Pa & 800 K, 20 kPa & 880 K, 30 kPa & 1800 K, 900 kPa \\
		\hline		
	\end{tabular}
	\label{tab:ablationTP}
	%\end{minipage}
\end{table*}

Even if the local conditions are sometimes insufficient for chemistry to occur at first, a necessary outcome of planetary growth is that the conditions change with time, namely: (a) previously too shallow deposits are increasingly buried within an externally growing atmosphere, and thus the conditions might be expected to turn in favor of chemistry at a later point; and (b) our chemical reactants are also generally expected to sink deeper into the atmosphere as a result of condensation. These aspects, in addition to possible convective mixing are discussed next.

\subsection{Compositional Stratification During Accretion}\label{SS:Stratification}
Once high-Z materials are present in the atmosphere, they are expected to form a compositional gradient inside the planet \citep{HelledStevenson-2017,LozovskyEtAl-2017,BrouwersEtAl-2018,BodenheimerEtAl-2018,VallettaHelled-2019,VazanHelled-2020,BrouwersEtAl-2020,OrmelEtAl-2021}. Indeed there is some seismic and gravity-based evidence for a deep compositional gradient in Saturn \citep{Fuller-2014,MankovichEtAl-2021} and Jupiter \citep{WahlEtAl-2017,DebrasChabrier-2019,HowardEtAl-2023}.

\begin{figure}[h!]
	%	\setcaptiontype{figure}
    \centering
	\begin{tabular}[b]{c}
		\includegraphics[scale=0.53]{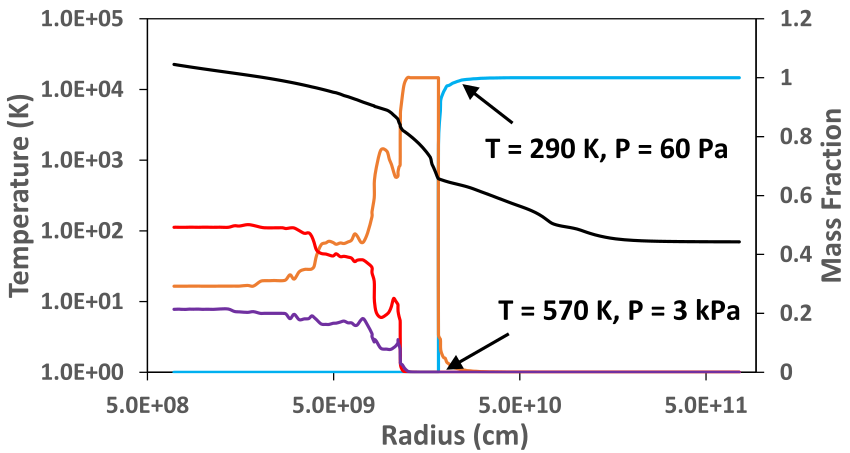}\label{fig:BodenheimerCode1_35}\\
		\small (a) 1.35 Myr
	\end{tabular}\\
	\begin{tabular}[b]{c}
		\includegraphics[scale=0.53]{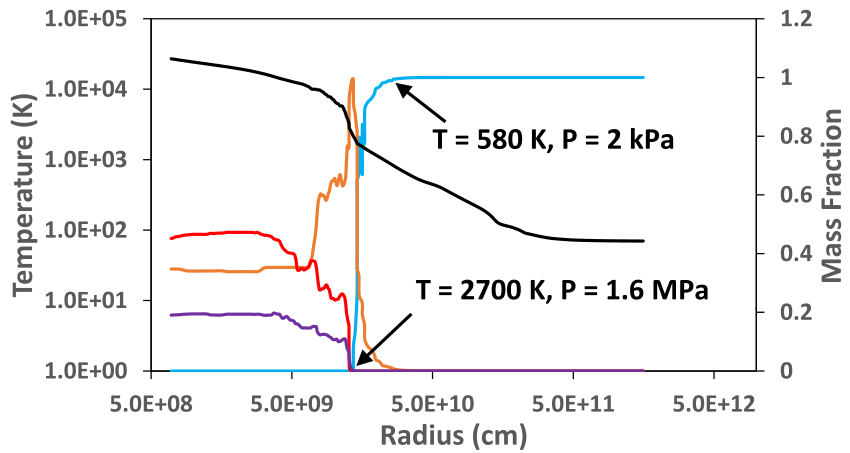}\label{fig:BodenheimerCode2_06}\\ 
		\small (b) 2.06 Myr
	\end{tabular}\\
	\begin{tabular}[b]{c}
		\includegraphics[scale=0.53]{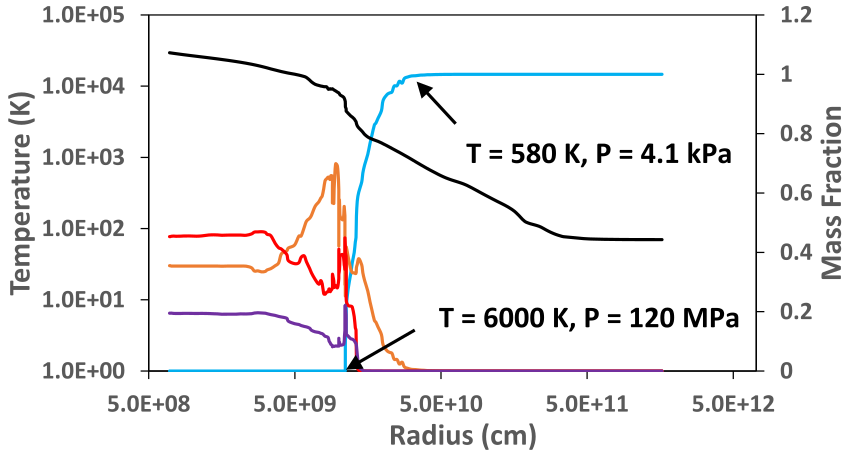}\label{fig:BodenheimerCode2_65}\\
		\small (c) 2.65 Myr
	\end{tabular}
	
	\caption{Growth of Uranus showing the mass fraction of iron (purple), silica (red), water (orange), and H+He (blue) on the right axis and the temperature (black) on the left axis, as a function of distance from the center. The arrows mark the T-P conditions where the composition changes from 100\% H+He to 0\% H+He. Panels (a)-(c) display this information at different times, providing the density, temperature and pressure profiles, required for later ablation and chemistry modeling.}
	\label{fig:BodenheimerCode}
\end{figure}

Assuming that high-Z vapor in the atmosphere obeys the ideal gas law, an assumption which was found to be judicious in giants such as Uranus and Neptune \citep{IaroslavitzPodolak-2007}, then compositional stratification can be explained as follows. At a given layer, and for a given high-Z material dissolved in the atmosphere by ablation, its local abundance determines its exact partial pressure in the gas. If the partial pressure of each high-Z material exceeds its saturated vapor pressure, which is a function of the ambient temperature in that layer, then this means that the high-Z material is in excess of a saturated state, and it must condense, releasing latent heat during the phase transition. Condensed matter descends towards the core if it is heavier than the ambient gas, losing gravitational energy on its way. This so-called 'rainout' process leads to a compositional gradient. Since the temperature increases with deepening layers, each layer is able to contain more vapor of the same material (the saturated vapor pressure increases with the temperature). At the boundary of the solid silicate core and the overlying saturated atmosphere, if the latter cannot absorb all the (continually) ablated high-Z material, the rest of it adds to the solid core. This allows the solid core to grow even without direct impacts \citep{IaroslavitzPodolak-2007,LeconteEtAl-2017,BodenheimerEtAl-2018,BrouwersEtAl-2018,BrouwersEtAl-2020}. The rainout process implies that high-Z materials increase inward according to Z. Given differences in their mean molecular weights, saturated vapor pressures and the temperature gradient in the atmosphere, one expects a stratified structure to emerge, just as it does in our growth and evolution code. The composition distribution for the fiducial materials iron, silica, water and H+He is shown in Figure \ref{fig:BodenheimerCode}.

Compositional gradients also provide the physical justification for our statistical random model code, which is currently implemented based on the idea of adjacent-material mixing and stratification. Convective mixing might in principle also hinder the rainout process and thus mix all of the various materials together. However, it has been shown previously that the formation of compositional gradients generally tend to prevent this from happening, stabilizing the layers against large-scale convective motion \citep{LeconteEtAl-2017,VazanHelled-2020}. Note that the rainout process will occur for any high-Z material, both the original materials introduced into the atmosphere by planetesimal ablation, or their chemical products, invoked by our chemical model below. Thus, even if the local conditions are initially incompatible for the chemistry we invoke, condensible species can still descend in the atmosphere until the required conditions are satisfied.

\subsection{Chemistry During Accretion}\label{SS:ChemistryModel}
Here we describe a simple chemistry model to facilitate the construction of an icy planet from ice-poor pebbles or planetesimals. In Section \ref{S:Intro} we have shown that accreted comet-like solids are mostly refractory, and in Section \ref{S:CometComposition} we have shown that 1/3-1/2 of these refractories are carbon-rich organics. By some estimates, aliphatic and aromatic carbons (consisting entirely of C and H) constitute more than 75\% by mass of the bulk organic matter \citep{DerenneRobert-2010} for meteroites, indicating that their mass is dominated by carbon. If we make a similar assumption for outer solar system refractories, graphite (pure C) might be taken to represent refractory organics in a simplified model. As additional validation of this assumption, we show in appendix \ref{A:KBOs} that KBOs and not merely comets, could have densities that are compatible with such a carbon-rich material. This implies that the solids which contributed material to Uranus and Neptune during their formation could be similar to the solids that were incorporated into the Kuiper belt, and therefore could be compatible with our model.

Our outer solar system planetesimals are represented by the following simplified components: Fe$_2$SiO$_4$ (fayalite) represents the rocky mineral component; C (graphite) represents the organic (or idealized carbon-rich) refractory component; The icy components are H$_2$O and CO. While there are many possible mineralogies to choose from, the one that was chosen above has a specific intention. By adding hydrogen, we can chemically separate Fe$_2$SiO$_4$ into Fe, SiO$_2$ and H$_2$O, while graphite will convert to CH$_4$, and CO to CH$_4$ and H$_2$O, and therefore all of these chemical products (iron,silica,water,methane and envelope) are standard materials that are most frequently used in canonical models of gas giants, and have well tested equations of state. We maintain flexibility in the model, by varying the ratios among the reactants.

We consider three idealized chemical reactions:

\begin{gather} \label{eq:fayalite}
	fayalite+hydrogen \rightarrow iron+silica+water\\
	Fe_2SiO_4+2H_2 = 2Fe+SiO_2+2H_2O \nonumber
\end{gather}

\begin{gather} \label{eq:graphite}
	graphite+hydrogen \rightarrow methane\\
	C+2H_2 = CH_4 \nonumber
\end{gather}

\begin{gather} \label{eq:carbonmonoxide}
	carbon~monoxide+hydrogen \rightarrow methane+water\\
	CO+3H_2 = CH_4+H_2O \nonumber
\end{gather}

We first discuss the relevance of these equations based on empirical evidence, from direct laboratory experiments. For the first reaction, some experiments show that the temperatures required in order to chemically decompose fayalite in the presence of molecular hydrogen, producing iron, silica and water, are in the range 550-670 K for pressures of 3.3 GPa and 1.4 GPa, respectively \citep{EfimchenkoEtAl-2019}. These authors showed that full decomposition takes 1-3 hours, and also that it requires a sufficient ratio of hydrogen to fayalite \citep{EfimchenkoEtAl-2021}. However, different experiments \citep{MassieonEtAl-1993} show that fayalite decomposition also occurs at far lower pressures, as long as the temperature is sufficiently high (e.g., 1070 K is required at a pressure of merely 1.4$\times 10^4$ Pa). A similar range of temperatures between 570-770 K and pressures spanning 0.5-3 GPa are likewise reported in lab experiments for our second chemical reaction, synthesizing methane from graphite and molecular hydrogen \citep{PenaAlvarez-2021}, although an exact reaction timescale is currently unavailable (private correspondence with authors). For our third reaction, methane and water are produced via hydrogenation of carbon monoxide, as known for more than a century \citep{SabatierSenderens-1902}. This process, also known as carbon monoxide methanation \citep{JohnsonEtAl-2015}, generally occurs in the lab at temperatures in excess of about 450 K and at modest pressures (around atmospheric pressure, or 1$\times 10^5$ Pa, but the reaction would likewise occur somewhat lower or higher pressures as well \citep{GaoEtAl-2012}, usually in the presence of various catalyzers). The idealized reaction timescale (without catalyzers) at a temperature of 777.5 K and pressure of 1.1$\times 10^4$ Pa is estimated at around one hour \citep{PodolakReynolds-1984}.

These laboratory experiments show that the original refractories invoked by our model are readily transformed through chemical reactions. However, the fayalite decomposition literature above demonstrated how a very wide range of temperature conditions led to the same experimental outcomes by greatly varying the pressure. This presents a problem since lab experiments are practically limited to a narrow set of conditions and can only partially probe the full parameter space of T-P conditions relevant to our problem. That being the case, we also investigated these reactions via theoretical chemical equilibrium calculations. Theory allows us to both validate experimental results, and also enable a full investigation of the entire range of T-P conditions relevant to Uranus and Neptune during their growth.

For chemical equilibrium calculations we utilize the widely used STANJAN code \citep{Reynolds-1986}, which provides the chemical equilibrium state of an ideal gas mixture. The list of elements that can be employed in STANJAN is quite extensive, however Fe is not one of the included elements in the code. Therefore, of our three reactions, only that of fayalite decomposition could not be accounted for with STANJAN. Fortunately, the fayalite decomposition experiments robustly cover a much wider range of T-P conditions compared to the other two reactions, spanning over 5 orders of magnitude in pressure \citep{MassieonEtAl-1993,EfimchenkoEtAl-2019}. As a first step we validate the two other experimental results for CO methanation \citep{GaoEtAl-2012}, and C methanation \citep{PenaAlvarez-2021}. We find that STANJAN reproduces the same results as the experiments, when we input the same T-P conditions into the code. We therefore gain confidence in its predictive capabilities, while we plan to use more elaborate codes in future studies.

In order to investigate methanation for the entire range of T-P conditions shown in Table \ref{tab:ablationTP}, we consider the elements C, H and O. Of course, since Fe and Si are excluded from STANJAN, the chemical equilibrium resulting in these calculations gives an incomplete picture, and is only designed to obtain a partial view of the chemistry. Using these elements we can account for the five species that participate in Equations \ref{eq:graphite} and \ref{eq:carbonmonoxide}, namely: H$_2$, C, CO, H$_2$O and CH$_4$. We require the relative fraction of each species in a state of equilibrium at constant pressure and temperature, but also for completion we look at several other potential chemical products which might stably form at the appropriate conditions: CO$_2$ (carbon dioxide), C$_2$H$_6$ (ethane) and CH$_3$OH (methanol). Since we must neglect fayalite, we assume as a heuristic approach, that our input planetesimals consist of twice as much C (40\% mass fraction) as H$_2$O (20\% mass fraction), and equal mass fraction of H$_2$O and CO (20\%). Our input mass fraction for H$_2$ (20\%) is selected arbitrarily in order to provide sufficient hydrogen for full chemical transformation of C, H$_2$O and CO. We note that our primary concern is not to explore various realizations of the input composition, but rather to investigate if the chemistry goes in the direction of methane or not. For this purpose, small changes in the initial composition do not qualitatively alter the outcomes. We demonstrate the effect of the temperature and pressure state on the chemistry in Table \ref{tab:STANJAN}, where we show the relative mass fractions of the various chemical products, given a range of conditions.

\begin{table*}[h!]
	\caption{The relative mass fractions of the products in our chemical reactions, given a range of temperature and pressure conditions experienced by dissociated high-Z materials during ablation, rainout or subsequent cooling of the planet. The initial mass fractions are shown in the parentheses.}
	\centering
	\smallskip
	\begin{tabular}{|c|c|c|c|c|c|c|c|c|c|}
		\hline
		\textbf{T-P state~\textbackslash~Species} & H$_2$ & C & CO & H$_2$O & CH$_4$ & CO$_2$ & C$_2$H$_6$ & CH$_3$OH \\  
		& (0.2) & (0.4) & (0.2) & (0.2) & & & & \\ \hline
		
		300 K, 100 Pa & 0.024 & 0 & 1E-17 & 0.329 & 0.648 & 3E-12 & 3E-12 & 8E-21 \\
		300 K, 10 kPa & 0.024 & 0 & 1E-21 & 0.329 & 0.648 & 3E-16 & 3E-12 & 8E-21 \\ \hline
		500 K, 1 kPa & 0.026 & 0 & 4E-05 & 0.319 & 0.643 & 0.012 & 1E-07 & 9E-13 \\
		500 K, 100 kPa & 0.024 & 0 & 5E-09 & 0.329 & 0.648 & 2E-06 & 1E-07 & 1E-12 \\ \hline
		750 K, 10 kPa & 0.087 & 0 & 0.148 & 0.093 & 0.5 & 0.171 & 6E-06 & 8E-10 \\
		750 K, 1 MPa & 0.025 & 0 & 1E-03 & 0.314 & 0.641 & 0.017 & 3E-05 & 1E-08 \\ \hline
		1000 K, 100 kPa & 0.129 & 4E-29 & 0.478 & 0.012 & 0.37 & 0.011 & 4E-05 & 6E-09 \\
		1000 K, 10 MPa & 0.04 & 4E-32 & 0.03 & 0.267 & 0.611 & 0.052 & 4E-04 & 7E-07 \\
		1000 K, 1 GPa & 0.024 & 0 & 2E-05 & 0.328 & 0.647 & 6E-05 & 7E-04 & 2E-06 \\ \hline
		2000 K, 1 MPa & 0.134 & 5E-10 & 0.511 & 2E-06 & 0.351 & 3E-07 & 4E-03 & 1E-09 \\
		2000 K, 100 MPa & 0.129 & 6E-14 & 0.485 & 0.015 & 0.364 & 2E-03 & 4E-03 & 1E-05 \\
		2000 K, 10 GPa & 0.036 & 6E-17 & 0.035 & 0.296 & 0.581 & 0.011 & 0.039 & 1E-03 \\ \hline
		6000 K, 100 MPa & 0.217 & 0.248 & 0.511 & 1E-04 & 0.023 & 5E-06 & 3E-04 & 5E-07 \\
		6000 K, 10 GPa & 0.129 & 4E-04 & 0.456 & 0.032 & 0.289 & 2E-03 & 0.089 & 3E-03 \\
		\hline		
	\end{tabular}
	\label{tab:STANJAN}
\end{table*}

We can divide the chemical processes that act on high-Z materials into three distinct stages in the evolution: initial ablation of planetesimals, their subsequent rainout and the long-term global evolution of the planet. For the first stage, we use our planetesimal ablation code. Thus, we consider the values listed in Table \ref{tab:ablationTP} to represent the relevant conditions. Cross referencing these values with Table \ref{tab:STANJAN}, it is clear that pebbles and small planetesimals are dissolved in the atmosphere early-on and experience conditions that drive the chemistry towards methane and water. Large planetesimals on the other hand might break up due to ram-pressure at locations where the conditions are more inclined to drive the chemistry towards less CH$_4$ and more CO (and even CO$_2$), while H$_2$O also diminishes. 

The second stage is rainout of high-Z materials towards the core. This process is included in our growth code for Uranus, however without explicitly accounting for chemistry, and only for mono-sized large planetesimals. We can nevertheless use it to approximately estimate the possible range in T-P conditions in the region of the atmosphere that contains hydrogen, as marked by the arrows in the various panel of Figure \ref{fig:BodenheimerCode}. Increase in both temperature and pressure during rainout may result in the chemistry moving either towards methane or carbon monoxide. It is hard to determine the outcome without including explicit feedbacks in the code.

The third stage occurs on long timescales and involves the global thermal evolution of the entire planet as it cools and contracts. We then expect a decrease in temperature and increase in pressure, overall. As an example, the present day pressures in the H+He outer envelope of Uranus and Neptune could be as high as a few hundred GPa (see Figure \ref{fig:CH4_internal_composition}) which is approximately three orders of magnitudes more than those obtained with our growth code for Uranus' early evolution (see Figure \ref{fig:BodenheimerCode}).

Table \ref{tab:STANJAN} generally shows that high temperature drives the chemistry towards carbon monoxide. However, keeping the same high temperature while increasing the pressure, always drives the chemistry back in the opposite direction, towards water and methane. This can be qualitatively understood from Le Chatelier's principle, which states that chemistry will move in the direction that undoes the pressure increase, i.e. in the direction of fewer molecules (the right hand side of Equation \ref{eq:carbonmonoxide}). Therefore, if we use our random statistical code to represent the more evolved state of the planet, we expect the chemistry to move towards methane.

Given the previous discussion, a full solution to the problem requires a complex numerical code which tracks the intricate thermal and physical evolution of the planet from early growth to the present day, while continually resolving the direction of chemistry both spatially and temporally. This code should also track the large scale convective motion in the interior in order to account for mixing or lack thereof. The chemical calculation should include additional elements and reactions. Such a code far exceeds the capabilities of any present-day study, and is also beyond the scope of this paper. While we cannot achieve the ultimate level of detail, here we use separate codes, showing that chemical methanation appears to be a likely outcome and is expected to occur. Future efforts must be taken in order to account for all the various feedbacks we discussed, in a self-consistent manner.

A last point to consider is the ratio of accreted hydrogen to high-Z material during the early evolution of the planet. If there is insufficient hydrogen, the chemical transformation will not go to completion. During the early stage of core accretion, models predict that the accreted material consists primarily of high-Z solids, and the relatively little hydrogen is probably insufficient to drive chemistry. At some point there is however a crossover where the bulk of the accreted material becomes dominated by H+He. This crossover depends on the details of the protoplanetary disc and whether accretion is dominated by km-sized solid bodies or by pebbles. For example, in a recent pebble accretion growth model \citep{OrmelEtAl-2021} this crossover will occur when a Uranus-mass planet has accreted about 1/3 of its final mass (however this work considers small semi-major axes). In our Uranus growth model the crossover point occurs when it has accreted about 2/3 of its mass (however our growth code does not consider pebble accretion). Thus, the assumption of full chemical transformation could break down, at least temporarily. Nevertheless, even if the planet has accreted much of its hydrogen during its final growth phase, the hydrogen could still interact with the interior of the planet if there is ongoing long-term mixing. In this case only the bulk fraction of hydrogen is important. Jupiter and Saturn are so massive that the bulk mass of H+He gas dwarfs the rest of the high-Z materials \citep{Helled-2019}. In Uranus and Neptune the H+He mass fraction is up to a few tens of \% \citep{HelledEtAl-2020}, which is enough to enable a full chemical transformation.

\subsection{Interior Composition After Methane Formation}\label{SS:ComparisonAfterMethanation}
As discussed above, the conditions are likely appropriate for chemistry to occur in the growing atmosphere of Uranus and Neptune. Demonstrating that chemical conversion is complete remains a future task for much more advanced models than currently exist in the literature. We note that even if the chemical transformation is incomplete, it still applies to a very large fraction of the accreted solids. Since modeling of partial chemical evolution is extremely difficult, we instead simplify the problem by assuming, without full justification, that the chemical reactions we invoke apply to all solids. The methane fraction resulting from the calculation below may thus be viewed as an upper limit only, since (a) we assumed complete chemical conversion; and (b) organic refractories are dominated by carbon, but are not 100\% carbon like we envisioned in our simple chemical scheme.

If the accreted material is indeed composed of comet-like solids, as we assumed, it is possible to estimate the effect of its methanation on the overall composition of the planet. Let $M_{\rm p}$ be the total mass of the planet, and $f_{\rm env}$ be the fraction of the accreted H+He envelope.  Let $r_{\rm h:he}$ denote the mass ratio of hydrogen to helium. The envelope mass will then be given by

\begin{equation}
	M_{\rm env} =  f_{\rm env}M_{\rm p}
\end{equation} 
and the mass of the high-Z materials is 
\begin{equation}
	M_{\rm Z} =  M_{\rm p} - M_{\rm env}.
\end{equation}

The masses of hydrogen and helium are, respectively,
\begin{equation}
	M_{\rm h} =  \frac{r_{\rm h:he}}{r_{\rm h:he}+1}M_{\rm env}
\end{equation}
and
\begin{equation}
	M_{\rm he} =  M_{\rm env} - M_{\rm h}.
\end{equation}

Let $r_{\rm ref:h2o}$ be the mass ratio of refractories to water in the accreted solids, $r_{\rm h2o:co}$ the mass ratio of water to CO, and $r_{\rm faya:c}$ the mass ratio of silicates (modeled by fayalite) to organic refractories (modeled by graphite, i.e. pure carbon).  The refractory/ice mass ratio in the planet's solid building blocks is then
\begin{equation}
	r_{\rm ref:ice} =  r_{\rm ref:h2o} \frac{r_{\rm h2o:co}}{r_{\rm h2o:co}+1}
\end{equation}
and the total refractory mass is 
\begin{equation}
	M_{\rm ref} =  \frac{r_{\rm ref:ice}}{r_{\rm ref:ice}+1}M_{\rm Z}.
\end{equation}

This can be divided between the fayalite (silicate) and graphite (carbon-rich organic) components, where
\begin{equation}
	M_{\rm faya} =  \frac{r_{\rm faya:c}}{r_{\rm faya:c}+1} M_{\rm ref}
\end{equation}
and
\begin{equation}
	M_{\rm c} =  M_{\rm ref} - M_{\rm faya}.
\end{equation}

Similarly, the ice mass is given by
\begin{equation}
	M_{\rm ice} =  \frac{1}{r_{\rm ref:ice}+1}M_{\rm Z}, 
\end{equation}
which can be split between water and carbon monoxide via
\begin{equation}
	M_{\rm h2o} =  \frac{r_{\rm h2o:co}}{r_{\rm h2o:co}+1}M_{\rm ice} 
\end{equation}
and
\begin{equation}
	M_{\rm co} =  M_{\rm ice} - M_{\textrm{h2o}}.
\end{equation}
 
%The molecular masses for these species are: m$_{\rm co}$=28$\times$amu, m$_{\rm h}$=1$\times$amu, m$_{\rm o}$=16$\times$amu, m$_{\rm h2}$=2$\times$m$_{\rm h}$, m$_{\rm h2o}$=m$_{\rm h2}$+m$_{\rm o}$, m$_{\rm c}$=12$\times$amu, m$_{\rm fe}$=56$\times$amu, m$_{\rm fe2}$=2$\times$m$_{\rm fe}$, m$_{\rm si}$=28$\times$amu, m$_{\rm o}$=16$\times$amu, m$_{\rm faya}$=m$_{\rm fe2}$+m$_{\rm si}$+4$\times$m$_{\rm o}$ and m$_{\rm sio2}$=m$_{\rm si}$+2$\times$m$_{\rm o}$, where amu is the atomic mass unit (1.66053907$\times 10^{-27}$ kg).

In what follows, let $N_{i}$ be the number of molecules of species $i$ and $m_{i}$ be the mass of such a molecule. By invoking chemistry we transform the above reactants, expecting the following reactions to occur:

\underline{Reaction 1}: fayalite converts to iron + silica (\ref{eq:fayalite}) -
This is a decomposition + single displacement type equation, where we assume that all fayalite accreted onto the planet decomposes into iron and silica, and displaces oxygen to make water from hydrogen.

The number of fayalite molecules is then
\begin{equation}
	N_{\rm faya} =  \frac{M_{\rm faya}}{m_{\rm faya}}.
\end{equation}

Reaction 1 thus reduces the total mass of hydrogen by 
\begin{equation}
	\Delta(M_{\rm h})_1 = - 2  m_{\rm h2} N_{\rm faya}
\end{equation}

and increases the mass of water by
\begin{equation}
	\Delta(M_{\rm h2o})_1 = \frac{m_{\rm o}}{m_{\rm faya}}M_{\rm faya} + 2m_{\rm h2}N_{\rm faya}
\end{equation}

The masses of iron and silica masses are given by
\begin{equation}
	M_{\rm fe} =  \frac{m_{\rm fe2}}{m_{\rm faya}}M_{\rm faya}
\end{equation}
and
\begin{equation}
	M_{\rm sio2} =  \frac{m_{\rm sio2}}{m_{\rm faya}}M_{\rm faya}.
\end{equation}
  
\underline{Reaction 2}: graphite converts to methane (\ref{eq:graphite}) -

This is a synthesis type equation, where we assume that all graphite accreted onto the planet combines with hydrogen to make methane. The number of graphite molecules formed is

\begin{equation}
	N_{\rm c} =  \frac{M_{\rm c}}{m_{\rm c}}
\end{equation}
and this decreases the hydrogen mass by
\begin{equation}
	\Delta(M_{\rm h})_2 = -2 m_{\rm h2} N_{\rm c}.
\end{equation}

The mass of methane produced is
\begin{equation}
	M_{\rm ch4} =  M_{\rm c} + 2 m_{\rm h2} N_{\rm c}.
\end{equation}

\underline{reaction 3}: all carbon monoxide converts to methane and water (\ref{eq:carbonmonoxide}) -

This is a double displacement type equation, where we assume that all carbon monoxide accreted onto the planet reacts with hydrogen, displacing oxygen and hydrogen to make methane and water. The number of carbon monoxide molecules is

\begin{equation}
	N_{\rm co} =  \frac{M_{\rm co}}{m_{\rm co}}.
\end{equation}

This reaction reduces the hydrogen mass by
\begin{equation}
	\Delta(M_{\rm h})_3 =   - 3 \times m_{\rm h2} \times N_{\rm co}.
\end{equation}

The methane mass increases by
\begin{equation}
	\Delta(M_{\rm ch4})_3 =  \frac{m_{\rm c}}{m_{\rm co}}M_{\rm co} + 2m_{\rm h2}N_{\rm co}
\end{equation}
and the water mass by
\begin{equation}
	\Delta(M_{\rm h2o})_3 =   \frac{m_{\rm o}}{m_{\rm co}}M_{\rm co} + m_{\rm h2} N_{\rm co}.
\end{equation}

We can now compare these estimates of the expected chemistry with the range of compositions determined by random interior modeling, noting that outcomes depend only on the ratios between the various reactants, and are independent on the planet mass. We therefore explore a range of values for the reactants, given the following parameters: (a) the planet's H+He envelope mass fraction $f_{\rm env}$. The maximum value is constrained by Fig. \ref{fig:CH4env_vs_watertorock} to be 0.28 and 0.18 for Uranus and Neptune, respectively. The minimum value is that which uses all the hydrogen in the atmosphere, enabling full chemical conversion of the accreted solids. It can be trivially calculated from the molecular weights of the reactants, giving a different result (in the range 0.1-0.17) for each combination of the free parameters, as specified in Table\,\ref{tab:modelMin}. There is the possibility of a still lower fraction, however without full chemistry. We do not consider this option here for simplicity, hence the resulting methane fractions we obtain may be viewed as upper limits; (b) the assumed mass ratio between H and He in the accreted gaseous material $r_{\rm h:he}$, which we take to be 3, i.e. close to solar; (c) the assumed mass ratio between refractories and water in the solid accreted material $r_{\rm ref:h2o}$. Based on the composition of Kuiper belt objects and comets, we take 2 and 4 as the lower and upper limits, respectively, with a mean of 3 (see Section \ref{S:Intro}); (d) the assumed mass ratio between water and carbon monoxide in the initial accreting ices, $r_{\rm h2o:co}$. We take 1 for the initially CO-rich end case, and 100 for the ratio in most contemporary comets, as another end case (see Section \ref{SS:CO_Models}); and (e) the assumed mass ratio between the original rocky (fayalite) and organic components (carbon) in the refractory materials $r_{\rm faya:c}$, which we take to be between 1 and 2 (see Section \ref{S:CometComposition}). 

The results of our simple chemical model are shown in Tables \ref{tab:modelMax} and \ref{tab:modelMin}, for the maximum and minimum H+He fractions, respectively (Table \ref{tab:modelMin} uses up all the molecular hydrogen in the atmosphere by construction, giving $f_{\rm h}=0$). The four left-hand columns depict the various initial choices for the free parameters. The initial ratio between H and He (solar) are kept fixed. The other six columns show the results of our chemical model in terms of the relative mass fraction of each material in the planet (noting that the results are independent on the planet's mass itself). Tables \ref{tab:modelMax} and \ref{tab:modelMin} unveil Uranus and Neptune as methane rich planets. Their methane mass fraction is in the range 20-50\%, comparable to and even larger than the mass fraction of water. However, here we considered full methanation of carbon-bearing species, which requires sufficiently massive H+He atmospheres, with a minimum mass fraction between 10-17\%.

\begin{table*}[h!]
	\caption{Chemical model results for maximum H+He fractions}
	\centering
	\smallskip
	%\begin{minipage}{14.2cm}
	\begin{tabular}{|l|l|l|l||l|l|l|l|l|l|}
		\hline
		$f_{\rm env}$ & $r_{\rm ref:h2o}$ & $r_{\rm h2o:co}$ & $r_{\rm faya:c}$ & $f_{\rm h}$ & $f_{\rm he}$ & $f_{\rm ch4}$ & $f_{\rm h2o}$ & $f_{\rm sio2}$ & $f_{\rm fe}$ \\ \hline
		\multicolumn{10}{|l|}{The maximum $f_{\rm env}$ is either 0.28 (Uranus) or 0.18 (Neptune); $r_{\rm h:he}$=3 is fixed.}\\ \hline
		
		0.28 & 2 & 1 & 1 & 0.1079 & 0.07 & 0.3429 & 0.3275 & 0.0529 & 0.0988\\
		0.18 & 2 & 1 & 1 & 0.0187 & 0.045 & 0.3905 & 0.373 & 0.0603 & 0.1125\\
		
		0.28 & 4 & 1 & 1 & 0.0996 & 0.07 & 0.3886 & 0.2395 & 0.0706 & 0.1318\\
		0.18 & 4 & 1 & 1 & 0.0092 & 0.045 & 0.4425 & 0.2728 & 0.0804 & 0.1501\\
		
		0.28 & 2 & 100 & 1 & 0.1251 & 0.07 & 0.3203 & 0.283 & 0.07 & 0.1313\\
		0.18 & 2 & 100 & 1 & 0.0383 & 0.045 & 0.3648 & 0.3223 & 0.0801 & 0.1496\\
		
		0.28 & 4 & 100 & 1 & 0.1082 & 0.07 & 0.3841 & 0.1954 & 0.0845 & 0.1578\\
		0.18 & 4 & 100 & 1 & 0.0191 & 0.045 & 0.4374 & 0.2225 & 0.0963 & 0.1797\\
		
		0.28 & 2 & 1 & 2 & 0.1267 & 0.07 & 0.2629 & 0.3381 & 0.0706 & 0.1318\\
		0.18 & 2 & 1 & 2 & 0.0402 & 0.045 & 0.2994 & 0.385 & 0.0804 & 0.15\\
		
		0.28 & 4 & 1 & 2 & 0.1247 & 0.07 & 0.2819 & 0.2536 & 0.0941 & 0.1757\\
		0.18 & 4 & 1 & 2 & 0.0378 & 0.045 & 0.3211 & 0.2888 & 0.1072 & 0.2001\\
		
		0.28 & 2 & 100 & 2 & 0.1501 & 0.07 & 0.214 & 0.297 & 0.0938 & 0.1751\\
		0.18 & 2 & 100 & 2 & 0.0668 & 0.045 & 0.2437 & 0.3383 & 0.1068 & 0.1994\\
		
		0.28 & 4 & 100 & 2 & 0.1383 & 0.07 & 0.2563 & 0.2123 & 0.1127 & 0.2104 \\
		0.18 & 4 & 100 & 2 & 0.0533 & 0.045 & 0.2919 & 0.2417 & 0.1284 & 0.2396\\ \hline
		
	\end{tabular}
	\label{tab:modelMax}
	%\end{minipage}
\end{table*}

\begin{table*}[h!]
	\caption{Chemical model results for minimal H+He fractions}
	\centering
	\smallskip
	%\begin{minipage}{14.2cm}
	\begin{tabular}{|l|l|l|l||l|l|l|l|l|l|}
		\hline
		$f_{\rm env}$ & $r_{\rm ref:h2o}$ & $r_{\rm h2o:co}$ & $r_{\rm faya:c}$ & $f_{\rm h}$ & $f_{\rm he}$ & $f_{\rm ch4}$ & $f_{\rm h2o}$ & $f_{\rm sio2}$ & $f_{\rm fe}$ \\ \hline		
		\multicolumn{10}{|l|}{The minimum $f_{\rm env}$ is calculated to give $f_{\rm h}=0$ at each combination; $r_{\rm h:he}$=3 is fixed.}\\ \hline
		
		0.159 & 2 & 1 & 1 & 0 & 0.0397 & 0.4005 & 0.3825 & 0.0618 & 0.1154\\		
		
		0.1698 & 4 & 1 & 1 & 0 & 0.0425 & 0.448 & 0.2762 & 0.0814 & 0.1519\\
		
		0.1359 & 2 & 100 & 1 & 0 & 0.0338 & 0.3844 & 0.3396 & 0.0844 & 0.1576\\
		
		0.1586 & 4 & 100 & 1 & 0 & 0.0225 & 0.4854 & 0.2469 & 0.1068 & 0.1994\\		
		
		0.1336 & 2 & 1 & 2 & 0 & 0.0397 & 0.4488 & 0.2283 & 0.0988 & 0.1844\\
		
		0.1364 & 4 & 1 & 2 & 0 & 0.0341 & 0.3381 & 0.3042 & 0.1129 & 0.2107\\
		
		0.0999 & 2 & 100 & 2 & 0 & 0.025 & 0.2675 & 0.3713 & 0.1173 & 0.2189\\
		
		0.1172 & 4 & 100 & 2 & 0 & 0.0293 & 0.3143 & 0.2603 & 0.1382 & 0.258\\ \hline
		
	\end{tabular}
	\label{tab:modelMin}
	%\end{minipage}
\end{table*}

\begin{figure}[h!]
	%	\setcaptiontype{figure}
	\begin{tabular}[b]{c}
		\includegraphics[scale=0.53]{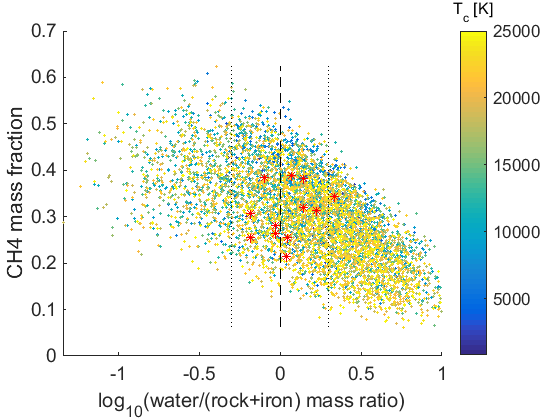}\label{fig:ch4_watertorock_CH4_Ura_22_withModels}\\
		\small (a) Uranus, MOI factor 0.22
	\end{tabular}
	\begin{tabular}[b]{c}
		\includegraphics[scale=0.53]{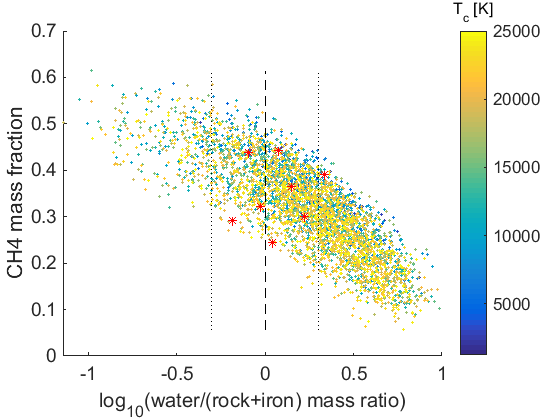}\label{fig:ch4_watertorock_CH4_Nep_24_withModels}\\ 
		\small (b) Neptune, MOI factor 0.24
	\end{tabular}
	
	\caption{The internal compositions resulting from chemistry are super-imposed (red asterisks) on the computer generated phase space of CH$_4$ mass fraction vs. water/rock mass ratio.}
	\label{fig:ch4_vs_watertorock_WithModels}
\end{figure}

In Figure \ref{fig:ch4_vs_watertorock_WithModels} all the models in Table \ref{tab:modelMax} (abundant hydrogen) are plotted to show the scatter in possible interior composition, against the background of randomly generated models. Using the most recent moment of inertia factor estimates for Uranus (0.22) and Neptune (0.24) \citep{NeuenschwanderHelled-2022} each random model is depicted by a single pixel, whose colour corresponds to the central temperature. The mass fraction of methane in the planet is shown as a function of the water/rock mass ratio (here rock = SiO$_2$+Fe). The dashed vertical line denotes equal proportions of water and rock. The right (left) dotted vertical line indicates a water/rock mass ratio of 2 (0.5). Superimposed on the same plot are the calculated upper-bound methane mass fractions based on chemistry, when accounting for the range in our reactants ratios (red asterisks). The composition derived from our simple chemical model falls well within the allowed parameter space of internal structures determined by our random models, independently producing the same patterns for both planets. In these models with abundant methane we see that the proportions between rock and water in Uranus and Neptune can shift either towards rock or water, but since methane itself is an ice, they are strictly always icy, despite having been initially formed by refractory-dominated solids. This is better reflected in the ice/rock mass ratio in Figure \ref{fig:CH4env_vs_icetorock} in Appendix \ref{A:CH4}. We likewise confirm that iron in our random models is within the expected range in solar system planets (Figure \ref{fig:CH4iron_vs_watertorock} in Appendix \ref{A:CH4}).

We note that the randomly generated models shown here assume solar composition for the H+He mixture (H/He$\sim$3), whereas models involving chemistry are always enriched in He (H/He$<$3). This inconsistency is however ignored for simplicity, since we cannot produce an infinite number of random models with every conceivable H/He ratio. We do however consider the end case scenario as well, where we produce random models that only contain He and no H. In this case too we find (not shown) that the methane fractions from Table \ref{tab:modelMin} fit the random models' scatter as well.

\section{Broader implications and predictions}\label{S:Implications}
Classical models of Uranus and Neptune present them as water-rich, yet their planetesimal building blocks should have a rock/water mass ratio greater than 2. We have explored the parameter space of planet models using a random planet generator and find that models fitting the mass, radius, and MOI factor of Uranus and Neptune can, at best, achieve a rock/water mass ratio somewhat less than 2, where the only alternative solution is to mix rock with a large fraction of hydrogen. We therefore suggest a different solution to this compositional tension problem: a significant fraction of the accreted planetesimals included highly carbon-rich refractory organic materials, similar to composition of comets. We show that while the oxygen in rocky minerals and small amounts of CO ice will react with the hydrogen to form additional water, the carbon inside refractory organics will form very significant amounts of methane. Our random model generator shows that such methane-rich planets can fit the observed properties of Uranus and Neptune. The rock/water mass ratio in these models now span a considerable range of values, but since a large fraction of methane is also present, we find that these models always contain more ice (methane and water) than rock and iron. Thus, the interior of 'ice giants' is indeed strictly icy.

Although the process described here would apply universally to other giant planets, for Jupiter and Saturn the fraction of high-Z material is negligible compared with that of H+He \citep{Helled-2019}, and, as a result, the CH$_4$ mass fraction expected will be too small to significantly affect the resulting models. Additionally, the interior pressures in these giants would easily dissociate most CH$_4$ into other hydrocarbons or carbon and hydrogen \citep{AncilottoEtAl-1997,BenedettiEtAl-1999,GaoEtAl-2010,ShermanEtAl-2012}. On the other hand, Uranus and Neptune, as well as other similar-sized exoplanets, are massive enough to accrete sufficient hydrogen in order to synthesize ample CH$_4$ but also just small enough to keep most of this methane from pressure dissociation. In addition, they do not accrete too much hydrogen and helium so that the heavier materials remain a significant fraction of the mass.
	
In addition to the fact that Uranus and Neptune might be methane-rich, there is another interesting outcome of our proposed chemistry. Since large amounts of atmospheric hydrogen are required for methanation, the atmospheric H/He ratio should be significantly sub-solar. We note that the Voyager atmospheric helium mass fraction for Uranus \citep{ConrathEtAl-1987} and for Neptune \citep{ConrathEtAl-1991} were actually found to be consistent with the primordial solar value. However, these measurements might be uncertain, noting that the Voyager values for the helium mass fraction in Jupiter have been revised upwards by the more recent measurements of Galileo \citep{VonzahnHunten-1996}, while the corresponding value for Saturn from Cassini differs by a factor of two \citep{AchterbergFlasar-2020}. Newer and more precise measurements of the helium mass fraction for Uranus and Neptune would therefore be of great interest. Additionally, it is important to note that the H/He ratio in the bulk of the atmosphere is difficult to measure using current remote techniques which only probe the uppermost atmosphere \citep{AtreyaEtAl-2020}. A Uranus mission has recently been recognized as a highest priority target by both NASA and ESA, and is expected to launch sometime during the next decade \citep{Fletcher2022,Mousis2022,Girija2023,Mandt2023}. We can expect an entry probe to reveal more high quality data about Uranus’ interior, especially if there is some mixing in the atmosphere, dredging up material from deeper within the planet.

\section{Conclusions}\label{S:Conclusions}
\begin{itemize}
    \item Astrophysical observations suggest that objects in the outer solar system, such as comets and Kuiper belt objects, are refractory-rich. Various pieces of evidence are reviewed (Section \ref{S:Intro}) which show that the mass ratio of refractory to icy materials in these objects is around 3, or higher. The outer planets Uranus and Neptune are, in contrast, referred to as ice giants, because classic models consider them to be water-rich. Hence, our paper formulates a problem - a tension exists between the composition of the outer planets and the planetesimals they likely accreted during their growth. We explore two possible solutions: either ice-poor planetesimals somehow prompted the formation of ice-rich planets, or the outer planets are actually rock-rich.
    \item We first review the refractory composition of outer solar system planetesimals, and find evidence in the literature that they contain a large mass fraction of refractory organics, between 1/3-1/2 (Section \ref{S:CometComposition}). This particular composition is rich in carbon, a fact which is later used to facilitate the first solution.
    \item At the same time, we investigate the detailed composition of Uranus and Neptune (Section \ref{S:UranusNeptuneComposition}). We use a computer code that generates hundreds of thousands of random density models for their interiors, commensurate with their observed radius, mass and moment of inertia. We formulate an algorithm that matches these random interior models with a particular composition. We find that planets made of iron, silica, water and hydrogen and helium envelope cannot have a rock/water mass ratio greater than 2, given our algorithm restriction of permitting mixing only between materials of adjacent molecular weights (Section \ref{SS:Water-rich}). Mathematically, mixing iron or silica with a large fraction of hydrogen might help overcome this limit, but it is not clear that this mixture is physically viable (Section \ref{SS:Rock-rich}). Adding carbon monoxide as a second ice does not aid in overcoming the aforementioned limit. We find that CO mostly replaces the rock and not the water (Section \ref{SS:CO_Models}). Adding methane as a second ice, however, leads to rock/water mass ratios in the range 0.1-10, because methane can replace water in large amounts. But the ensuing methane mass fraction in the random planet interiors, between 10-50\%, is very challenging to explain (Section \ref{SS:CH4_models}).
    \item  We conclude that the solution of a rock-rich interior for Uranus and Neptune is possible, but it requires a fine-tuned state of material mixing that might not be feasible. We suggest instead that the abundant organic-rich refractories in outer solar system planetesimals are key to a more natural solution, producing copious amounts of methane through chemical reactions in the hydrogen-rich environment of the growing outer planets (Section \ref{S:Growth}). 
    \item Planetesimals or pebbles are accreted into the proto-planet's atmosphere and facilitate its growth, while dissolving in the atmosphere and forming compositional gradients based on the material molecular weights (Section \ref{SS:Stratification}). A simplified chemical model is described to facilitate the conversion of C-rich materials into methane (Section \ref{SS:ChemistryModel}). We compare the models obtained after this chemical methanation, considering a range of free parameters, to our random interior models with methane, and find an excellent agreement (Section \ref{SS:ComparisonAfterMethanation}).
    \item A methane-rich interior therefore seems to present a natural solution to the composition tension problem of the outer planets. We find that Uranus and Neptune are massive enough to accrete sufficient hydrogen, allowing the proposed chemistry, but unlike Jupiter and Saturn, they are not so massive that methane will fully pressure-dissociate in the interior. Additionally, the sum of all high-Z materials inside the major gas giants is dwarfed by their large hydrogen and helium fraction. Uranus, Neptune and similar exoplanets are therefore just in the sweet spot that enables methane-rich planets. Because hydrogen is consumed through chemistry, the bulk He/H ratio is predicted to be larger than solar. We propose this as a priority task for future exploration missions of the outer planets (Section \ref{S:Implications}).

\end{itemize}

\section{CRediT authorship contribution statement}
\textbf{Uri Malamud:} Conceptualization, Methodology, Software, Formal Analysis, Investigation, Resources, Data curation, Writing - original draft, Writing - review \& editing, Visualization, Project administration, Funding acquisition. \textbf{Morris Podolak:} Software, Formal analysis, Investigation, Writing - review \& editing, Funding acquisition. \textbf{Joshua Podolak:} Software. \textbf{Peter Bodenheimer:} Software.

%UM: initiated and oversaw the project; suggested the composition tension problem and the proposed solution that chemistry of comet-like refractory materials produces methane-rich planets; took part in designing the statistical code, ran it and analyzed its results; analyzed the results of the STANJAN chemical equilibrium code; performed the chemical conversion calculations and the other analytical calculations; wrote the main text and appendix sections. MP suggested the chemical transformation of carbon monoxide to methane; initiated the statistical code project; wrote and ran the planetesimal ablation code and analyzed its results; suggested using the STANJAN chemical equilibrium code; contributed to the preparation of the manuscript, the figures and the analysis. JP programmed the statistical code and contributed to its design. PB contributed Uranus' growth and evolution code.

\section{Declaration of competing interest}
The authors declare that they have no known competing financial
interests or personal relationships that could have appeared to influence the work reported in this paper.

\section{Data availability}
The data that has been used is available on request from the corresponding author.

\section{Acknowledgements}
We wish to thank Marc Brouwers for useful conversations with regard to giant planet growth and evolution, Vadim Efimchenko for providing information about his fayalite decomposition laboratory experiments and Eugene Gregoryanz for providing information about his laboratory experiments regarding the methanation of graphite/diamond. UM acknowledges support by the Niedersächsisches Vorab in the framework of the research cooperation between Israel and Lower Saxony under grant ZN 3630 and grant by MOST-space. MP acknowledges support by a grant from the Pazy Fund of the Israel Atomic Energy Commission. PB acknowledges support from NASA grant EW-18-2-0060.

\newpage

\appendix

\section{Are large Kuiper belt objects compatible with comet-like composition?}\label{A:KBOs}\
\renewcommand{\thefigure}{A\arabic{figure}}
\setcounter{figure}{0}
In the main text we discuss how comet-like refractory material can be composed of a large fraction of organic rich material, up to 50\% in mass. It needs to be shown that such organic-rich refractory constituents are also compatible with the contemporary bulk density of intermediate- and large-sized KBOs.

An underlying assumption in this work, as well as other studies, is that objects in the outer solar system form roughly in the same region and therefore have similar compositional characteristics \citep{KenyonEtAl-2008,MalamudPrialnik-2015,BiersonNimmo-2019}. KBOs are expected to incorporate materials similar to those in comets, and Uranus and Neptune are likewise hypothesized (as the main rationale for this work) to incorporate similar materials as these two classes of objects - a premise which gave rise to the composition tension problem this paper addresses. Unlike comets, however, KBOs are unique in their ability to test this hypothesis, and therefore we will show that comet-like, organic-rich material, is compatible with the composition in the Kuiper belt.

We start our discussion with intermediate-sized KBOs, whose internal structure contains non-negligible residual porosity. We then discuss Pluto, the most well-characterized object in the Kuiper belt which is both massive and sufficiently evolved in order to suppress internal porosity.

A decade ago, a trend among KBOs was identified -- an increase in the bulk density of dwarf planets as a function of their mass \citep{Brown-2012}. The composition, or rock/water mass ratio which was commonly suggested in the past for Kuiper belt dwarf planets, was based on the casual assumption that porosity is negligible. If porosity is negligible, then the bulk density translates directly into composition. If however porosity is non-negligible, then it implies that the derived rock/water mass ratio is only a lower limit.

A true determination of the composition from the bulk density, must take into consideration how porosity is internally distributed within the dwarf planet as a function of the internal (evolving) pressure and temperature profiles, as well as grain texture, and mixing (if rock and ice are mixed or differentiated). To this purpose, a differentiation model for icy bodies \citep{MalamudPrialnik-2013} was revised \citep{MalamudPrialnik-2015} to include an equation of state suitable for porous icy bodies with radii of a few hundred km, based on available empirical studies of ice and rock compaction. The plausibility of this equation of state was then tested on three intermediate-sized KBOs of various masses, considering how these icy bodies evolve due to internal heating. The porosity profile evolves as a result of internal migration of water, rock/water stratification and compaction. It was shown that the present-day bulk densities in the model were quite well matched with the observed trend, without requiring one to vary the initial composition of those three objects.

Based on this aforementioned work \citep{MalamudPrialnik-2015}, it was concluded that the trend is well produced for a rock/water mass ratio of approximately 3, and a similar conclusion was later echoed in a different study \citep{BiersonNimmo-2019}. The exact rock/water mass ratio predicted by such models, depends on what one assumes for the rock specific density (i.e. its grain density). According to the former study \citep{MalamudPrialnik-2015}, two phases of rocky material were considered. The rock was initially considered to be anhydrous, as in comets \citep{WoodenEtAl-2017}. Later interaction by the flow of liquid water, which was explicitly followed in the code, transformed the rock and rendered it into the hydrated phase. The specific density of the hydrated phase is typically about 20\% lower than the anhydrous phase \citep{MalamudPrialnik-2013}. The anhydrous rock specific density was a free parameter. Values were considered in the range 3250-3500 kg m$^{-3}$. A rock/water mass ratio of 3 matched the Kuiper belt density trend, corresponding to the 3250 kg m$^{-3}$ value. Slightly changing the value to 3000 or 3500 kg m$^{-3}$, only modestly changed these results.

According to the above arguments, for the conclusions to apply, it is merely necessary to justify their choice of rock specific density. Taking the mineralogy assumed in the present work, that is, a binary combination of fayalite and graphite, we can calculate what the specific grain density should be. At zero porosity, as well as low pressure and temperature which would be typical of dwarf-planet sized objects, fayalite and graphite have the specific densities $\varrho_{\rm faya}$=4390 kg m$^{-3}$ and $\varrho_{\rm c}$=2260 kg m$^{-3}$, respectively. Assuming an additive volume law, the density of their combined refractory mixture $\rho_{\rm ref}$ is given by 

\begin{equation}\label{eq:RefDensity}
	\rho_{\rm ref} =  \left(\frac{x_{\rm faya}}{\varrho_{\rm faya}}+\frac{x_{\rm c}}{\varrho_{\rm c}}\right)^{-1}=\left(\frac{1-x_{\rm c}}{\varrho_{\rm faya}}+\frac{x_{\rm c}}{\varrho_{\rm c}}\right)^{-1},
\end{equation} 

where $x_{\rm faya}$ and $x_{\rm c}$ are the respective mass fractions of fayalite and graphite in the mixture, and $x_{\rm faya} + x_{\rm c} = 1$.  From Equation \ref{eq:RefDensity}, a 50:50 ratio leads to a density of 3000 kg m$^{-3}$, and a 63:37 ratio leads to 3250 kg m$^{-3}$. These mix ratios, where organic fractions are in the range 37-50\%, are in excellent agreement with those proposed in the Section \ref{S:CometComposition} for outer solar system material. The grain specific density $\varrho_{\rm ref}$ of anhydrous refractories is therefore compatible with previous work \citep{MalamudPrialnik-2015}, and will lead to the assumed rock-rich intermediate-sized KBOs, having a rock/water mass ratio of around 3.

Turning to Pluto for additional clues, its mass is much larger than the previously mentioned intermediate-sized KBOs. The pressure exerted by its own self-gravity and the higher internal temperatures attained during its evolution are likely sufficient in order to have rendered its porosity close to zero \citep{BiersonEtAl-2018}. It is therefore possible to directly derive Pluto's composition without a great deal of porosity-related degeneracy, adopting a few simplifying assumptions, as follows. 

Let us assume that the porosity is either exactly zero, or allow for the possibility of some small porosity $\psi$ which should be retained primarily in the icy mantle, where the pressures are estimated in the range 0-200 MPa between the surface and the core-mantle boundary \citep{HussmannEtAl-2006}, and yield porosities between 0-30\% \citep{MalamudPrialnik-2015}. Let us further consider a two-layered stratified structure - a rocky core underlying an icy mantle. For simplicity, water is the only ice accounted for. We also ignore the fact that differentiation leads to a more intricate structure in both core and mantle, rather than a simple two-layer scheme (ignoring partial hydration and dehydration in the core \citep{MalamudEtAl-2017}). Finally, let us assume that the core contains the original refractories which accreted to form Pluto (i.e. ignoring hydration), and the mantle consists of pure water ice. In that case Pluto's density is given by

\begin{equation}\label{eq:PlutoDensity}
	\rho_{\rm pluto} =  \left(\frac{x_{\rm ref}}{\rho_{\rm ref}}+\frac{x_{\rm h2o}}{\varrho_{\rm h2o}(1-\psi)}\right)^{-1}=\left(\frac{x_{\rm ref}}{\rho_{\rm ref}}+\frac{1-x_{\rm ref}}{\varrho_{\rm h2o}(1-\psi)}\right)^{-1},
\end{equation} 

where $\varrho_{\rm h2o}$=917 kg m$^{-3}$ is the specific density of water ice, $x_{\rm ref}$ and $x_{\rm h2o}$ are the respective mass fractions of the original refractories and water in the dwarf planet, and $x_{\rm ref} + x_{\rm h2o} = 1$. The residual porosity in the mantle $\psi$ can be zero, or it can be estimated as some relatively small number. We take 15\% as an average value between 0 and 30\% discussed above. We now substitute $\rho_{\rm ref}$, combining Equations \ref{eq:RefDensity} and \ref{eq:PlutoDensity}, in order to obtain the value of $x_{\rm c}$ as a function of the other parameters, leading to

\begin{equation}\label{eq:GraphiteFraction}
	x_{\rm c} =  \left( \frac{1}{x_{\rm ref}} \left( \frac{1}{\rho_{\rm pluto}}-\frac{1}{\varrho_{\rm h2o}(1-\psi)}\right) + \left( \frac{1}{\varrho_{\rm h2o}(1-\psi)}-\frac{1}{\varrho_{\rm faya}}\right) \right) \cdot \left( \frac{1}{\varrho_{\rm c}}-\frac{1}{\varrho_{\rm faya}}\right)^{-1}.
\end{equation} 

\begin{figure}[h!]
%	\setcaptiontype{figure}
	\begin{tabular}[b]{c}
		\includegraphics[scale=0.57]{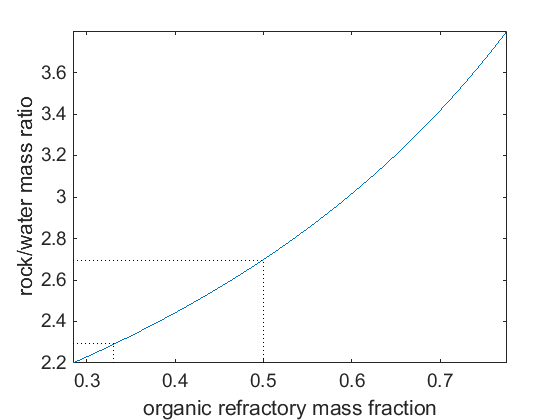}\label{fig:psi0}\\
		\small (a) $\psi$=0
	\end{tabular}
	\begin{tabular}[b]{c}
		\includegraphics[scale=0.57]{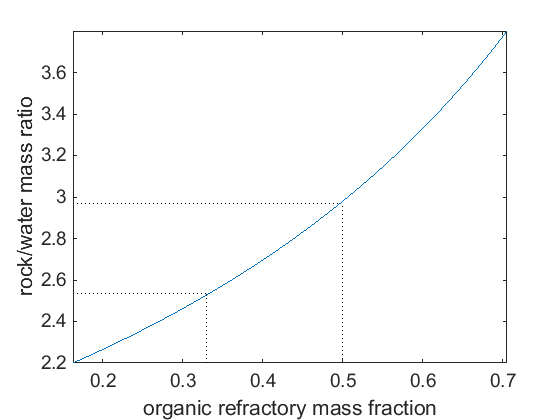}\label{fig:psi0_05}\\ 
		\small (b) $\psi$=0.05
	\end{tabular}
	\begin{tabular}[b]{c}
		\includegraphics[scale=0.57]{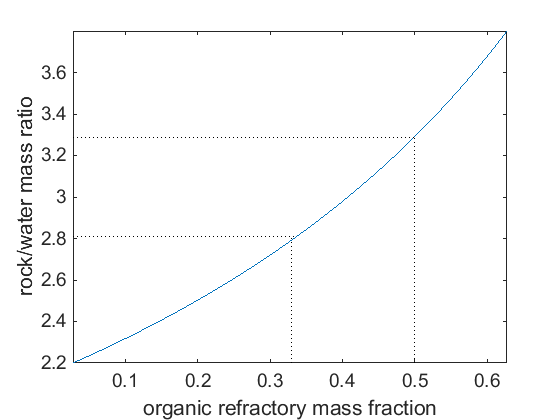}\label{fig:psi0_1}\\
		\small (c) $\psi$=0.1
	\end{tabular}
	\begin{tabular}[b]{c}
		\includegraphics[scale=0.57]{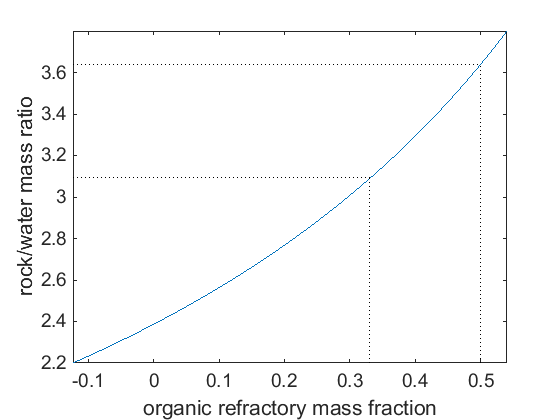}\label{fig:psi0_15}\\
		\small (d) $\psi$=0.15
	\end{tabular}

	\caption{Pluto's rock/water mass ratio as a function of the organic mass fraction in refractory material, for different mantle porosities $\psi$. The dotted lines correspond to the plausible range of organics, based on measurements from comet Halley (see SI).}
	\label{fig:r_vs_Xc}
\end{figure}

Letting $\rho_{pluto}$=1854 kg m$^{-3}$ as determined from observations \citep{SternEtAl-2015}, and replacing the refractory fraction $x_{\rm ref}$ with the rock/water mass ratio $r$ using the relation $x_{\rm ref}=r/(1+r)$, we can use the above relation in order to evaluate $r$ as a function of $x_{\rm c}$. Figure \ref{fig:r_vs_Xc} shows four subplots for different choices of mantle residual $\psi$, up to 15\%.

Although we have made several simplifying assumptions, the well measured bulk density of Pluto helps to constrain its composition using the analytic formulation above. The dotted lines in Figure \ref{fig:r_vs_Xc} correspond to the expected range in organic mass fraction. The lower and upper limits on the x-axis are based on the direct mass-spectrometer measurements obtained during the Giotto mission to comet Halley (as was discussed in the main text). It can be seen that Pluto's rock/water mass ratio falls in the range 2.4-3.6. While there is some uncertainty in the parameters, these calculations are commensurate with the rock-rich composition of Pluto, and organic-rich refractories may be invoked for its constituent building blocks. Both Pluto and intermediate-sized KBOs support this conclusion.

\clearpage

\section{Mission Stardust inability to provide bulk carbon content in comet dust}
\label{A:Stardust}
\renewcommand{\thefigure}{B\arabic{figure}}
\setcounter{figure}{0}
In the main text we introduced several arguments supporting a large fraction of carbon-rich organics in cometary dust. Here we add a note about why the Stardust mission cannot be used in this context, despite being the only comet sample return mission to date. 

The Stardust spacecraft rendezvoused with comet 81P/Wild 2 at a distance of 1.86 AU, and using an aerogel collector, a low-density highly porous silica-based material, captured grains that were released from the nucleus. Encased within a capsule, the sample returned to Earth for analysis \citep{Brownlee-2014}. 

Although stardust has accomplished an unprecedented achievement so far in the study of comets, and has delivered a wealth of information, it is less relevant for uncovering the exact fraction and nature of organics on its host comet, based on the arguments in \cite{ClemettEtAl-2010} and references therein.

First, the capture in the aerogel during the flyby occurred at a velocity of around 6 km s$^{-1}$. The kinetic energy of a particle impacting aerogel at this speed is dominated by the linear kinetic energy term. For organic grains in particular, this energy typically exceeds their vaporization energy by an order of magnitude. Heating must therefore affect (remove or at least alter) cometary organics to a considerable degree during capture. Additionally, the aerogel itself incorporates organic contaminates originating in its manufacture. Since they are impossible to remove completely, this presents a problem in the sample analysis. Finally, the grains which are collected by Stardust spent an estimated 20 minutes to several hours in the coma prior to collection, which is ample time for an organic grain, or at least a significant fraction of it, to dissociate by interacting with stellar photons.

Since the heat of capture is most severe at the entry site into the aerogel, a few Stardust organic particles which piggy-backed behind terminal tracks made by previously collected large particles, were able to survive to be analyzed \citep{deGregorioEtAl-2011}. Nevertheless, our overall judgment based on the aforementioned arguments is that the Stardust mission cannot reliably provide an assessment of the bulk organic fraction.

\clearpage

\section{Additional information for models with methane}\label{A:CH4}
\renewcommand{\thefigure}{C\arabic{figure}}
\setcounter{figure}{0}
The data shown in Figures \ref{fig:CH4env_vs_watertorock} and \ref{fig:ch4_vs_watertorock} was used in the main text, depicting some statistical properties related to models where CH$_4$ was used as a second ice. Here we expand on some additional properties of the random models.

\begin{figure}[h!]
	%	\setcaptiontype{figure}
	\begin{tabular}[b]{c}
		\includegraphics[scale=0.53]{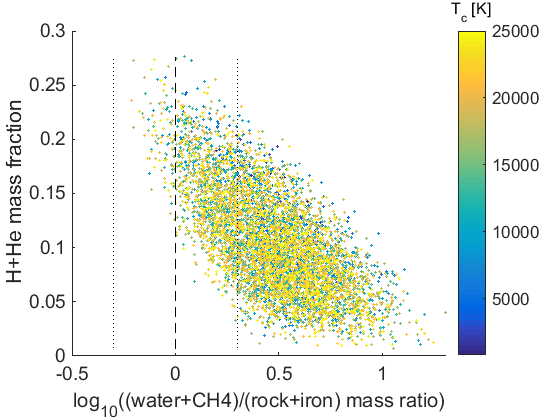}\label{fig:env_icetorock_CH4_Ura_22}\\
		\small (a) Uranus, MOI factor 0.22
	\end{tabular}
	\begin{tabular}[b]{c}
		\includegraphics[scale=0.53]{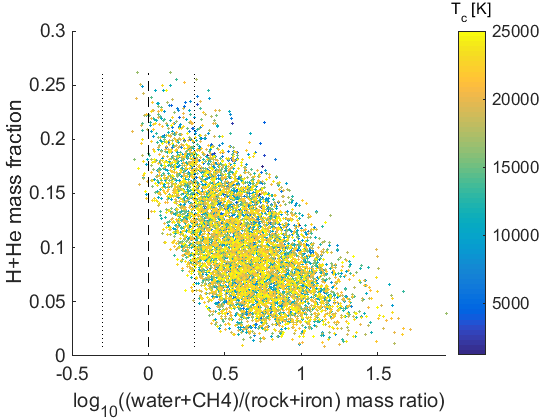}\label{fig:env_icetorock_CH4_Ura_23}\\ 
		\small (b) Uranus, MOI factor 0.23
	\end{tabular}
	\begin{tabular}[b]{c}
		\includegraphics[scale=0.53]{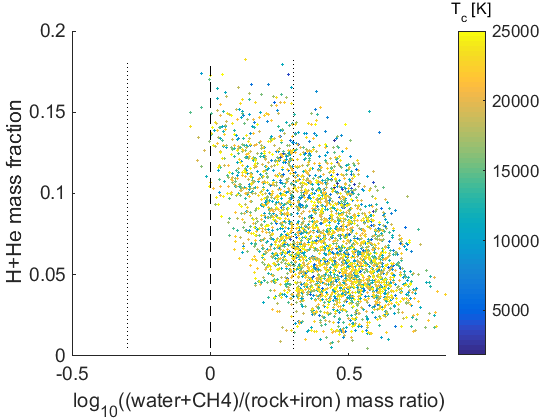}\label{fig:env_icetorock_CH4_Nep_23}\\
		\small (c) Neptune, MOI factor 0.23
	\end{tabular}
	\begin{tabular}[b]{c}
		\includegraphics[scale=0.53]{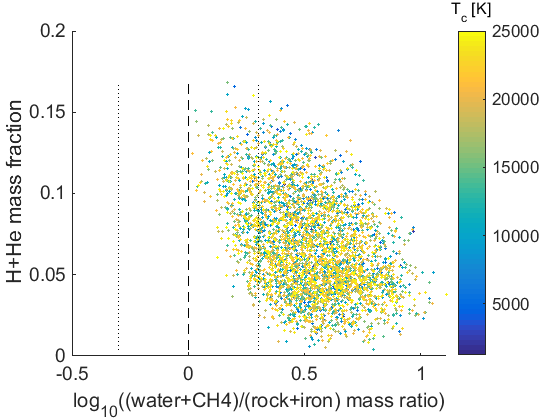}\label{fig:env_icetorock_CH4_Nep_24}\\
		\small (d) Neptune, MOI factor 0.24
	\end{tabular}
	
	\caption{Mass fraction of H+He vs. ice/rock mass ratio (logarithmic). Each point denotes a random interior model, color coded by its central temperature. The dashed vertical line indicates equal fractions of water and rock. The right (left) dotted vertical line indicates a ice/rock mass ratio of 2 (0.5).}
	\label{fig:CH4env_vs_icetorock}
\end{figure}

In Figure \ref{fig:CH4env_vs_icetorock} we show that the ice/rock mass ratio is larger than the water/rock mass ratio in Figure \ref{fig:CH4env_vs_watertorock}. This is however trivial, since the numerator (ice) now includes an additional icy material (methane) and is therefore larger than water alone. The limit ice/rock mass ratio of Uranus and Neptune is between 0.5-1 (or rock/ice between 1-2, equivalently).

\begin{figure}
	%	\setcaptiontype{figure}
	\begin{tabular}[b]{c}
		\includegraphics[scale=0.53]{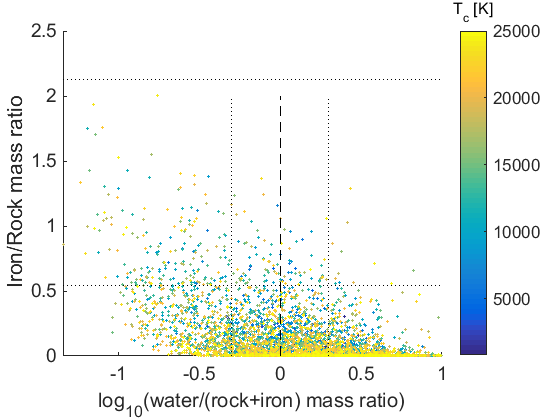}\label{fig:iron_watertorock_CH4_Ura_22}\\
		\small (a) Uranus, MOI factor 0.22
	\end{tabular}
	\begin{tabular}[b]{c}
		\includegraphics[scale=0.53]{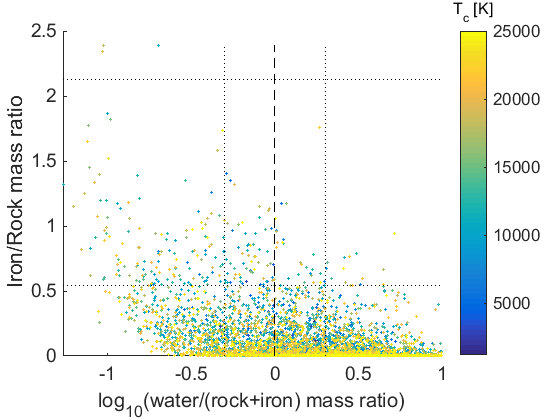}\label{fig:iron_watertorock_CH4_Ura_23}\\ 
		\small (b) Uranus, MOI factor 0.23
	\end{tabular}
	\begin{tabular}[b]{c}
		\includegraphics[scale=0.53]{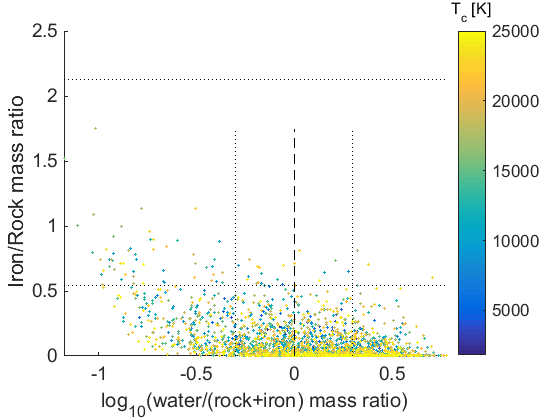}\label{fig:iron_watertorock_CH4_Nep_23}\\
		\small (c) Neptune, MOI factor 0.23
	\end{tabular}
	\begin{tabular}[b]{c}
		\includegraphics[scale=0.53]{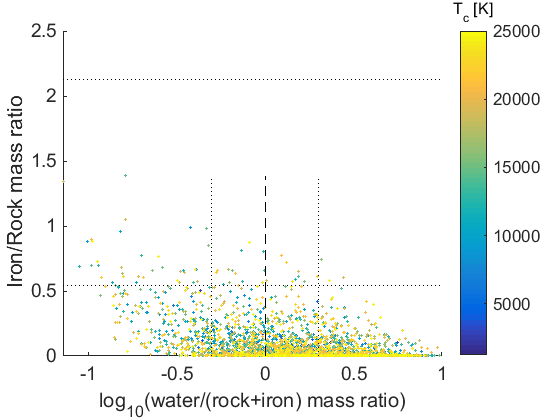}\label{fig:iron_watertorock_CH4_Nep_24}\\
		\small (d) Neptune, MOI factor 0.24
	\end{tabular}
	
	\caption{iron/rock mass ratio vs. water/rock mass ratio (logarithmic). Each point denotes a random interior model, color coded by its central temperature. The dashed vertical line indicates equal fractions of water and rock. The right (left) dotted vertical line indicates a water/rock mass ratio of 2 (0.5). The lower (upper) horizontal line indicates the iron/rock mass ratio of Earth (Mercury).}
	\label{fig:CH4iron_vs_watertorock}
\end{figure}

Figure \ref{fig:CH4iron_vs_watertorock} shows the iron/rock mass ratio 
on the y-axis and the water/rock mass ratio on the logarithmic x-axis. We truncate the water/rock mass ratio at a value of 10. Here we test the feasibility of random compositions in terms of their iron fraction. In order to judge if the iron fraction is not too high, we compare the iron/rock mass ratio of each model to that of the known ratio in solar system planets. Two horizontal dash-dotted lines are drawn. The bottom line refers to the ratio on Earth, and the top line to Mercury \citep{ReynoldsSummers-1969}. Mercury is considered an upper limit since it probably has the highest iron fraction among the planets, perhaps due to a mantle-striping impact in its history \citep{BentzEtAl-1988}. Models above the top line are therefore viewed as non-physical. We however see that the majority of models fall below the bottom line or between the bottom and top lines, in broad agreement with expectation, noting that to our knowledge there is no firm lower bound to the expected iron content.

\clearpage
\bibliographystyle{icarus2}
\bibliography{bibfile}

\end{document}